\documentclass{article}
\usepackage{bm}
\usepackage[utf8]{inputenc}
\usepackage{amsmath}
\usepackage{cancel}
\usepackage{graphicx}
\usepackage{color}

\usepackage{comment}
\usepackage{ amssymb }
\usepackage{soul}

\newcommand{\beginsupplement}{%
        \setcounter{table}{0}
        \renewcommand{\thetable}{S\arabic{table}}%
        \setcounter{figure}{0}
        \renewcommand{\thefigure}{S\arabic{figure}}%
        \setcounter{section}{0}
        \renewcommand{\thefigure}{S\arabic{figure}}%
     }

\usepackage{geometry}
 \newcommand{\Ipred}{I_{\rm pred}}
 \newcommand{\Ipast}{I_{\rm past}}

\title{Predicting a noisy signal: the costs and benefits of time averaging as a noise mitigation strategy}
\author{Jenny Poulton, Age Tjalma, Lotte Slim, Pieter Rein ten Wolde}

\begin{document}
\maketitle

\begin{abstract}
One major challenge for living cells is the measurement and prediction of signals corrupted by noise. In general, cells need to make decisions based on their compressed representation of noisy, time-varying signals.  Strategies for signal noise mitigation are often tackled using Wiener filtering theory, but this theory cannot account for systems that have limited resources and hence must compress the signal. To study how accurately linear systems can predict noisy, time-varying signals in the presence of a compression constraint, we extend the information bottleneck method. We show that the optimal integration kernel reduces to the Wiener filter in the absence of a compression constraint. This kernel combines a delta function at short times and an exponential function that decays on a timescale that sets the integration time. Moreover, there exists an optimal integration time, which arises from a trade-off between time averaging signal noise and dynamical error. As the level of compression is increased, time averaging becomes harder, and as a result the optimal integration time decreases and the delta peak increases. We compare the behaviour of the optimal system with that of a canonical motif in cell signalling, the push-pull network, finding that the system reacts to signal noise and compression in a similar way.
\end{abstract}

\section{Introduction}

Autonomous or self-perpetuating systems such as cells typically exist in dynamic environments. A general requirement for self-perpetuating systems to thrive in such environments is the ability to respond to changing conditions. Ideally, a system would make an instantaneous change to respond to an environmental change. In reality, mounting a response takes time. Given this, an optimal response requires systems to predict an environmental change~\cite{sachdeva2021optimal, malaguti2021theory}. Intriguingly, experiments have revealed that even single-celled organisms can predict environmental change~\cite{mitchell2009adaptive,tagkopoulos2008predictive}. For example, cells can use the arrival of one sugar to predict that the next one will arrive~\cite{mitchell2009adaptive}. In this work, we consider the optimal prediction of time-varying signals. We consider biological sensing systems, but our ideas can be applied to any system predicting a time-varying signal.

Living cells live in rich sensory environments and can sense and react to many different external signals. These include light, motion and chemical concentrations. In this work, we consider the trajectory of the changing concentration of ligand molecules in the environment as a function of time. These concentrations are measured via receptors, which are typically located on the surface of the cell. The ligand molecules bind to these receptors, which transmit information to a downstream system within the cell. Receptor-ligand binding, like all processes at the cellular scale, is noisy. As a result, the signal that is propagated to the downstream system is corrupted by signal noise, also called input noise. Living cells, like any signal detection system, are thus inevitably affected by signal noise. 
This work is interested in understanding how systems can mitigate the effect of this signal noise.

 How cells can maximize their sensing precision by minimizing the propagation of signal noise has been studied extensively. In their pioneering paper, Berg and Purcell~\cite{berg1977physics} pointed out that cells can reduce the sensing error via the mechanism of time integration. In this mechanism, the cell does not infer the ligand concentration from the current concentration but rather from its average over some given integration time. Following the work of Berg and Purcell, many studies have addressed the question of how accurately living cells can measure ligand concentrations via the mechanism of time integration \cite{Bialek2005,Wang2007,Rappel2008,govern2012fundamental,Mehta2012,govern2014,Govern2014b,Kaizu2014,Fancher:2017ba}. Importantly, these studies assume that the signal is constant on the timescale of the response and that the different signal values are averaged uniformly in time.  However, when the integration time is comparable to the correlation time of receptor-ligand binding, the optimal weighting becomes non-uniform \cite{govern2012fundamental}. Moreover, ligand concentrations often fluctuate on a timescale that is comparable to the response time of the system, as, for example, in chemotaxis \cite{Tostevin2009,2021.Mattingly}. Predicting these signals optimally requires a non-uniform time average  \cite{Hinczewski:2014iq,becker2015optimal}. Another sensing strategy, which can reach a higher sensing precision, is that of maximum-likelihood sensing \cite{mora2010limits,Lang:2014ir} or Bayesian filtering \cite{Mora:2019fd}.

Since systems cannot generally respond instantaneously to changes in their environment, it becomes beneficial to anticipate the change and mount a response ahead of time. How accurately this can be done is determined not by how accurately they can predict the current signal but rather the future signal. For a system to predict the future, it  must extract characteristics of the past signal that are informative about the future signal. The amount of predictive information stored in the past signal trajectory about the future signal is the mutual information \(I(\vec{\bf{s}},s(\tau))\) between the past signal trajectory $\vec{\bf s}$ and the signal value  $s(\tau)$ at a future timepoint $\tau$. This predictive information puts a fundamental lower bound on the prediction error. However, signal noise means that this bound can, in general, not be reached. 
Wiener filtering theory \cite{wiener1949extrapolation,kolmogorov1962interpolation, ShannonWiener} makes it possible to derive, for linear systems, the optimal integration function that minimizes the prediction error for time-varying  signals in the presence of signal noise, and it has been applied to cellular systems \cite{Hinczewski:2014iq,becker2015optimal}.


Wiener filtering theory, however, does not recognize that systems are built with finite resources. In general, and as assumed in Wiener filtering theory, systems do not predict the future signal $s(\tau)$ from the input signal trajectory $\vec{\bf s}$ directly, but rather indirectly, from the output of the signalling system, $x$. This output depends on the past input signal trajectory $\vec{\bf s}$. Wiener filtering theory assumes that the input trajectory can be reliably mapped onto the output \(x\).
In general, however, the output trajectory $x(t)$ is a noisy and compressed representation of the input trajectory $s(t)$ because resources such as protein copy numbers and energy are finite. The data processing inequality implies that the mutual information between the compressed output $x$ and the future of the signal $s(\tau)$ is less than that between the uncompressed signal and the future: \(I(x;s(\tau))\leq I(\vec{\bf{s}},s(\tau))\). In this work, we go beyond Wiener filtering theory to study systems which have limited resources.

Here, we study the optimal compression of the input signal into the output for prediction under resource constraints. We define the optimal compression as that which maximises the predictive information in the compressed output \(\Ipred=I(x;s(\tau))\) 
subject to the constraint that the information the output has about the past, \(\Ipast=I(x;\vec{\bf{s}})\), is limited. We will confine ourselves to linear systems, and derive the optimal compression via the information bottleneck method~\cite{Tishby1999,chechik2003information}.  

The information bottleneck method has been applied to a wide range of biological systems. The method has led to a greater understanding of optimal gene expression patterns for fly development and has identified the optimal sensors associated with this process~\cite{bauer2021trading,bauer2022does}. It has been used to analyse retinal ganglion cells~\cite{sachdeva2021optimal,palmer2015predictive}, finding that the retina provides a nearly optimal representation of the predictive information contained in the visual input~\cite{palmer2015predictive}. A related work calculates whether position or velocity information is more useful to the retina for predicting a moving image~\cite{sachdeva2021optimal}. Yet, none of these studies has directly considered signals that are corrupted by signal noise. In this work, we will extend the Gaussian information bottleneck presented by Tishby et al.~\cite{chechik2003information} to systems with signal noise. Using this approach, we will derive the optimal integration function, which captures the characteristics of the past signal that are most informative about the future signal in the presence of signal noise and a compression constraint.

In section \ref{InfBot}, we will outline the discrete information bottleneck for systems with Gaussian signal noise. This method combines the information bottleneck method and the Wiener filter, considering both signal noise and compression. Previous attempts to link the information bottleneck method and the Wiener filter have not included signal noise on which the kernel acts~\cite{Matz}, without which the Wiener filter does not straightforwardly apply.

In section \ref{AR}, we introduce a discrete Markovian signal modelled with an autoregressive model of order \(1\). A Markovian signal is the simplest signal in which the past is predictive of the future. We will then add correlated Gaussian noise to that signal, also modelled with an autoregressive model of order \(1\).  

In section \ref{KernelsSect}, we  address the optimal prediction of this signal in the presence of signal noise and resource constraints. We  derive optimal kernels for compressing the past signal and calculate the amount of predictive information these compressed representations contain. We find that the optimal kernel combines a \(\delta\) peak at zero with a decaying exponential, which allows for time averaging over the signal noise. The relative importance of these two contributions, as well as the integration time (the timescale on which the exponential contribution decays), depends on the compression level. When the  resources are limited, and the compression level is high, the \(\delta\) peak is relatively large, and the integration time is short because the system cannot time average. In the other limit, the system time averages over an optimal integration time, which arises from the interplay between time averaging and the dynamical error or signal distortion \cite{Hinczewski:2014iq,becker2015optimal,malaguti2021theory}. Additionally, the relative contribution of the \(\delta\) peak reduces. Finally, we examine the effect of changing the variance and the correlation time of the noise. When the noise variance is larger,  more priority is given to the exponential part of the kernel, and its range, the integration time, also increases because this allows for more time averaging. When the correlation time of the noise is larger, the exponential part of the kernel widens to enable effective time averaging, while the importance of the \(\delta\) peak increases because time averaging becomes less successful. 

In section \ref{pushpullsect}, we will compare our optimal kernels with the kernel of a well-known biological signalling motif, the push-pull network \cite{Goldbeter1981}.  Push-pull systems are omnipresent in prokaryotic and eukaryotic cells \cite{Alon:2007tz}. Examples are phosphorylation cycles, as in MAPK cascades; GTPase cycles, such as the Ras system; and two-component systems, including the chemotaxis system of Escherichia coli. Push–pull networks constitute a simple exponential filter \cite{Hinczewski:2014iq,becker2015optimal,malaguti2021theory}, and hence do not contain a contribution with a $\delta$ peak implying that the push-pull motif is not optimally compressing the signal.

This work develops a very general method which can be used to study optimal compression for prediction in noisy systems with resource constraints. While we use it to study biological systems, the effect of compression on systems predicting any number of noisy signals, from financial data to robotic sensing data, can be studied using this method.

\section{Deriving the information bottleneck for a system with signal noise}\label{InfBot}

This work seeks to find the optimal scheme for compressing a signal to predict the future given constrained resources. To start, we must define a general process that captures the essence of the problem, and that can be optimised. We consider signals that obey Gaussian statistics, which are corrupted by noise that also obeys Gaussian statistics. It has been shown that the optimal response systems for these signals are linear\cite{chechik2003information}. We, therefore, consider systems that respond linearly over the range of input fluctuations:
\begin{equation}
    x=\vec{\bf{A}}(\vec{\bf{s}}+\vec{\bf{\eta}})+\xi.
    \label{output}
\end{equation}
Here \(\vec{\bf{s}}\) is a vector representing a discretised signal trajectory, \(\vec{\bf{\eta}}\) is a vector representing a noise trajectory. The linear kernel \(\vec{\bf{A}}\) is a vector and \(\vec{\bf{A}}(\vec{\bf{s}}+\vec{\bf{\eta}})\) is a scalar representing a weighted average over all timepoints of the signal corrupted by signal noise. The compression noise \(\xi\) is a scalar. Thus our compressed output \(x\) is a scalar. This output is correlated with the value of the signal in the future, and we are interested in its correlation with the value at one particular timepoint \(\tau\) into the future, the scalar \(s(\tau)\).

The information bottleneck finds the optimal kernel \(\vec{\bf{A}}\) over the signal trajectory to maximise predictive information while also compressing the signal. This optimal compression is found by maximising the information bottleneck Lagrangian with respect to \(\vec{\bf{A}}\):
\begin{equation}
    \max_{\vec{\bf{A}}}\mathcal{L}=\max_{\vec{\bf{A}}}\left(\Ipred-\gamma \Ipast\right).\label{Lagrangian2}
\end{equation}
Here \(\gamma\) is a Lagrange multiplier that dictates the compression level. When this Lagrangian is maximised, the mutual information between the compressed output and the signal value in the future is maximised subject to the constraint that the mutual information between the compressed output and the signal trajectory in the past is limited.  
The compression level \(\gamma\) runs from zero to one. Recall that \(x\) is a compression of \(\vec{\bf{s}}\), so \(\Ipred\leq\Ipast\). Given this, at \(\gamma=1\) optimal \(\Ipred=\Ipast=0\). At lower \(\gamma\), the system is allowed to increase \(\Ipast\) via \(\vec{\bf{A}}\) to make a better prediction of the future. We can rewrite the information bottleneck Lagrangian in terms of entropy using \(I(a;b)=H(a)-H(a|b)\). The entropy for a Gaussian system in one dimension is \(H(\Sigma_{x})=\frac{1}{2}\log{|\Sigma_{x}|}\) where \(\Sigma_{x}\) is the covariance of \(x\) (where \(x\) is a vector, \(\Sigma_{x}\) is a covariance matrix). The conditional entropy \(H(\Sigma_{y|z})=\frac{1}{2}\log{|\Sigma_{y|z}}|\) where \(\Sigma_{y|z}\) is the conditional covariance of variable \(y\) given variable \(z\) where one or both of \(y\) and \(z\) can be vectors. Combining these, we rewrite the information bottleneck Lagrangian as
\begin{align}
    \max_{\vec{\bf{A}}}\mathcal{L}&=H(x)-H(x|s(\tau))-\gamma H(x)+\gamma H(x|\vec{\bf{s}})\\
    \max_{\vec{\bf{A}}}\mathcal{L}&=(1-\gamma)\frac{1}{2}\log{|\Sigma_{x}|}-\frac{1}{2}\log{|\Sigma_{x|s(\tau)}|}+\gamma\frac{1}{2}\log{|\Sigma_{x|\vec{\bf{s}}}|}.\label{Lagrangian}
\end{align}
The information bottleneck absent signal noise eliminates \(x\) from \(\Sigma_{x}\), \(\Sigma_{x|\vec{\bf{s}}}\) and \(\Sigma_{x|s(\tau)}\), then differentiates with respect to \(\vec{\bf{A}}\), resulting in an eigenvalue equation~\cite{chechik2003information}.

We follow the same method, but because of the addition of signal noise, the definition of the covariances has changed. Given \(x=\vec{\bf{A}}(\vec{\bf{s}}+\vec{\bf{\eta}})+\xi\) and noting that here there are no correlations between the signal \(\vec{\bf s}\), the signal noise \(\vec{\bf \eta}\), and the compression noise \(\xi\), respectively, we find that:
\begin{align}
    \Sigma_{x}&=\langle \delta x \delta x\rangle\\
    &=\langle \delta (\vec{\bf{A}}(\vec{\bf{s}}+\vec{\bf{\eta}})+\xi) \delta ((\vec{\bf{s}}^T+\vec{\bf{\eta}}^T)\vec{\bf{A}}^T+\xi)\rangle\\
    &=\vec{\bf{A}}\langle \delta \vec{\bf{s}} \delta \vec{\bf{s}}^T\rangle \vec{\bf{A}}^T + \vec{\bf{A}}\langle \delta \vec{\bf{\eta}} \delta \vec{\bf{\eta}}^T \rangle \vec{\bf{A}}^T +\langle \delta \xi \delta \xi \rangle\\
     &=\vec{\bf{A}}\Sigma_{\vec{\bf{s}}}\vec{\bf{A}}^T + \vec{\bf{A}}\Sigma_{\vec{\bf{\eta}}}\vec{\bf{A}}^T +\Sigma_{\xi}
\end{align}
If \(\vec{\bf{s}}\) is known, the remaining uncertainty in \(x\) is \(\vec{\bf{A}}\vec{\bf{\eta}}+\xi\). Hence, \(\Sigma_{x|\vec{\bf{s}}}=\vec{\bf{A}}\Sigma_{\vec{\bf{\eta}}}\vec{\bf{A}}^T+\Sigma_{\xi}\). Finally, to find \(\Sigma_{x|s(\tau)}\) we use the Schur complement formula: \(\Sigma_{x|s(\tau)}=\Sigma_{x}-\Sigma_{x s(\tau)}\Sigma_{s(\tau)}^{-1}\Sigma_{s(\tau)x}\). Now \(\Sigma_{xs(\tau)}=\langle\delta x\delta s(\tau) \rangle=\vec{\bf{A}}\langle\delta \vec{\bf{s}}\delta s(\tau)\rangle=\vec{\bf{A}}\Sigma_{\vec{\bf{s}} s(\tau)}\) and similarly \(\Sigma_{s(\tau)x}=\Sigma_{s(\tau)\vec{\bf{s}}}\vec{\bf{A}}^T\). Thus
\begin{align}    
\Sigma_{x|s(\tau)}&=\vec{\bf{A}}\Sigma_{\vec{\bf{s}}}\vec{\bf{A}}^T+\vec{\bf{A}}\Sigma_{\vec{\bf{\eta}}}\vec{\bf{A}}^T+\Sigma_{\xi}-\vec{\bf{A}}\Sigma_{\vec{\bf{s}} s(\tau)}\Sigma_{s(\tau)}^{-1}\Sigma_{s(\tau) \vec{\bf{s}}}\vec{\bf{A}}^T\\
    &=\vec{\bf{A}}\left(\Sigma_{\vec{\bf{s}}}-\Sigma_{\vec{\bf{s}} s(\tau)}\Sigma_{s(\tau)}^{-1}\Sigma_{s(\tau) \vec{\bf{s}}}\right)\vec{\bf{A}}^T+\vec{\bf{A}}\Sigma_{\vec{\bf{\eta}}}\vec{\bf{A}}^T+\Sigma_{\xi}\\
    &=\vec{\bf{A}}\Sigma_{\vec{\bf{s}}|s(\tau)}\vec{\bf{A}}^T+\vec{\bf{A}}\Sigma_{\vec{\bf{\eta}}}\vec{\bf{A}}^T+\Sigma_{\xi}.
\end{align}
The information bottleneck Lagrangian (equation \ref{Lagrangian}) can be rewritten as
\begin{align}
    \max_{\vec{\bf{A}}}{\mathcal{L}}&=(1-\gamma)\frac{1}{2}\log{(\vec{\bf{A}}\Sigma_{\vec{\bf{s}}}\vec{\bf{A}}^T+\vec{\bf{A}}\Sigma_{\vec{\bf{\eta}}}\vec{\bf{A}}^T+\Sigma_{\xi})}+\gamma\frac{1}{2}\log{(\vec{\bf{A}}\Sigma_{\vec{\bf{\eta}}}\vec{\bf{A}}^T+\Sigma_{\xi})}\\\nonumber
    &-\frac{1}{2}\log{(\vec{\bf{A}}\Sigma_{\vec{\bf{s}}|s(\tau)}\vec{\bf{A}}^T+\vec{\bf{A}}\Sigma_{\vec{\bf{\eta}}}\vec{\bf{A}}^T+\Sigma_{\xi})}.
\end{align}
Differentiating and setting equal to zero gives
\begin{align}
    \frac{d\mathcal{L}}{dA}&=(1-\gamma)\frac{\vec{\bf{A}}(\Sigma_{\vec{\bf{s}}}+\Sigma_{\vec{\bf{\eta}}})}{\vec{\bf{A}}(\Sigma_{\vec{\bf{s}}}+\Sigma_{\vec{\bf{\eta}}})\vec{\bf{A}}^T+\Sigma_{\xi}}+\gamma\frac{\vec{\bf{A}}\Sigma_{\vec{\bf{\eta}}}}{\vec{\bf{A}}\Sigma_{\vec{\bf{\eta}}}\vec{\bf{A}}^T+\Sigma_{\xi}}\\\nonumber
    &-\frac{\vec{\bf{A}}(\Sigma_{\vec{\bf{s}}|s(\tau)}+\Sigma_{\vec{\bf{\eta}}})}{\vec{\bf{A}}\Sigma_{\vec{\bf{s}}|s(\tau)}\vec{\bf{A}}^T+\vec{\bf{A}}\Sigma_{\vec{\bf{\eta}}}\vec{\bf{A}}^T+\Sigma_{\xi}}=0.
\end{align}
Unlike the system without signal noise~\cite{chechik2003information}, this equation no longer reduces to an eigenvalue equation and must be solved numerically.

Can this method be compared to the Wiener filter? The Wiener filter minimises the mean squared error between the filter output (here \(x\)) and the signal at a present or future time (here, the signal at a future point \(s(\tau)\)). For a Gaussian system, minimising the mean squared error, \(|\left(x^2-s(\tau)\right)^2|\), is equivalent to maximising the mutual information, \(\Ipred\). Maximising this mutual information is equivalent to maximising the information bottleneck Lagrangian, \(\mathcal{L}=\Ipred-\gamma \Ipast\), for \(\gamma=0\). Thus, as \(\gamma\rightarrow0\), the optimal kernels found by the information bottleneck method converge to that which optimally filters out signal noise, given by the Wiener filter. Convergence to the Wiener filter will generally be true for kernels found using this method. This explicit link has only been made possible by including signal noise on which the kernel acts, a vital component of the Wiener filter problem. Some attempt has been made to link the information bottleneck method and the Wiener filter before~\cite{Matz}. However, this work fails to include a noise source acted on by the kernel. Since the Wiener filter traditionally mitigates noise via the kernel, this rendered this comparison between the IBM and the Wiener filter somewhat confusing. 

\section{A discrete signal modelled by an autoregressive model}\label{AR}

Since our method of calculating the information bottleneck is discrete, we need a discrete signal. We consider a discrete Markovian input signal given by an autoregressive model. We choose a Markovian process as the simplest example of a signal in which the future depends on the past and can therefore be predicted. The autoregressive model is a time-series model, where each value is regressed upon previous values in sequence~\cite{cryer2008time}. This work will focus on an order \(1\) autoregressive model with a zero mean, which models a Markovian process: 
\begin{equation}
    S_t= \phi_1 S_{t-1} + \sigma_{AR}^2\eta.
\end{equation}
Here \(\phi_1\) is the weighting of how element \(1\) in the past affects the current value, \(\eta\) is a white noise process of variance one and mean zero and \(\sigma^2_{AR}\) sets the full variance of the white noise term. The covariance function of an order \(1\) autoregressive model with zero mean is
\begin{equation}
    \langle\delta_S(0)\delta_S(t)\rangle=\frac{\sigma_{AR}^2\phi_1^{\frac{t}{dt}}}{1-\phi_1^2}.
\end{equation}
Here the current time \(t\) must be an integer multiple of the timestep \(dt\). At non-integer multiples of \(dt\), the function is not defined. Where the function is defined, we want this discrete covariance function to take the same form as that for continuous Markovian signal with covariance function \(\langle\delta_S(t_{1})\delta_S(t_{2})\rangle=\sigma_{s}^2 e^{-\frac{1}{\tau_{s}} |t_{1} -t_2|}\). Here \(\sigma_{s}^2\) is the variance of the signal, and \(\tau_{s}\) is the correlation time of the signal. To give the autoregressive function the same correlation function, we take \(\phi_1 = e^{-\frac{dt}{\tau_s}}\) and \(\sigma_{AR}^2=\sigma_s^2\left(1-e^{-\frac{2dt}{\tau_s}}\right)\).

Our signal is corrupted by correlated signaling noise with covariance \(\langle\delta_{\vec{\bf{\eta}}}(t_{1})\delta_{\vec{\bf{\eta}}}(t_{2})\rangle=\sigma_{\eta}^2 e^{-\frac{1}{\tau_\eta}|t_{1} -t_2|}\), where \(\sigma_{\eta}^2\) is the variance of the noise and \(\tau_\eta\) is the correlation time of the noise. This is once again modelled by an autoregressive function with \(\phi_1 = e^{-\frac{dt}{\tau_\eta}}\) and \(\sigma_{AR}^2=\sigma_\eta^2\left(1-e^{-\frac{2dt}{\tau_\eta}}\right)\).

\section{Optimal kernels and the information bottleneck limits}\label{KernelsSect}
\begin{figure}
    \centering
    \includegraphics[scale=0.4]{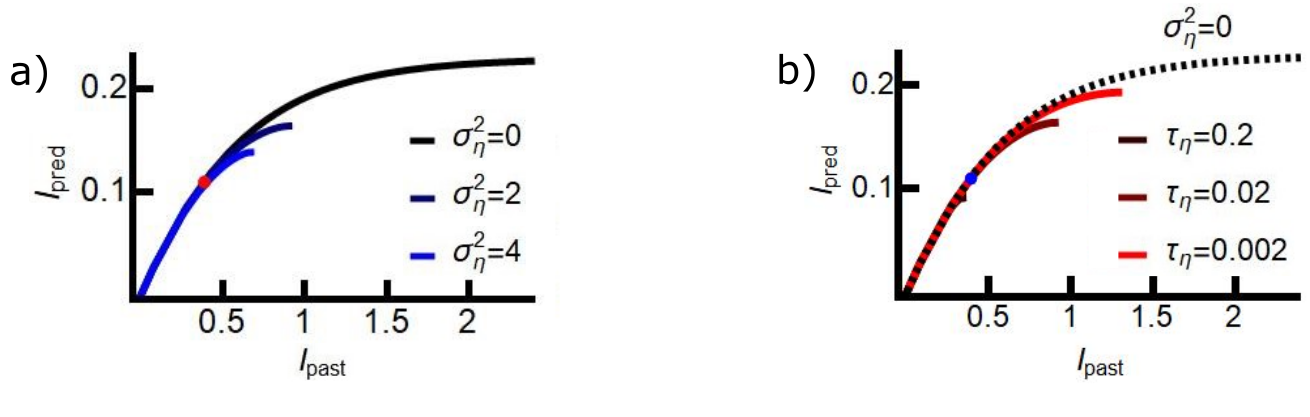}
    \caption{In both graphs, \(\sigma_s^2=1\) and \(\tau_s=2\)s. {\bf As the noise variance \(\sigma_{\eta}^2\) (a) or noise correlation time \(\tau_{\eta}\) (b) increases, the maximum amount of \(\Ipred\) and \(\Ipast\) available to the system decreases.} We calculate optimal kernels at fixed \(\tau_\eta\) and \(\sigma_\eta^2\) for varying \(\gamma\). These kernels result in a given \(\Ipred\) and \(\Ipast\), which we plot parametrically. \(\gamma\) is zero for the top right part of each curve and increases towards one at the bottom left part of the curve. The curve for \(\sigma_\eta^2=0\) is identical to that found using the information bottleneck without signal noise developed in ref. ~\cite{chechik2003information}. This limit is the maximum amount of predictive information per bit of past information that can be extracted from a given signal trajectory in the absence of signal noise. When signal noise is present, the system can extract less predictive information from the trajectory, and this effect increases as the variance of the noise increases (a). We also note that the maximum amount of past information about the trajectory decreases as \(\sigma^2_\eta\) increases. The amount of information about the past and the present also decreases as the correlation time of the noise increases (b). This is because a kernel requires a longer integration time to mitigate correlated noise, which increases dynamical error. In a) \(\tau_\eta=0.02\)s, in b) \(\sigma_\eta^2=2\), in both, \(\sigma_s^2=1\), \(\tau_s=2\), \(\sigma_\xi^2=1\), \(\tau=1\)s. \(dt=0.01\)s and \(T=0.5\)s. Dots mark optimal kernels from fig. \ref{Optimum}.}
    \label{infcurves}
\end{figure}

\begin{figure}
    \centering
    \includegraphics[scale=0.4]{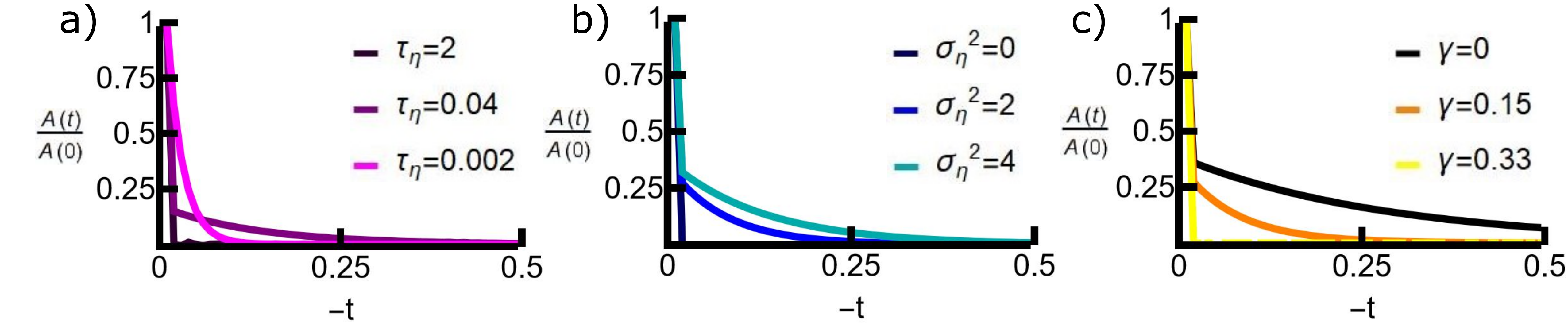}
    \caption{We plot optimal kernels for various \(\sigma_\eta^2\), \(\tau_\eta\) and \(\gamma\). \(\gamma\) defines the level of compression. Low \(\gamma\) corresponds to the high \(\Ipred\) and \(\Ipast\) region and high \(\gamma\) corresponds to the low \(\Ipred\) and \(\Ipast\) region. The kernels are discrete, but we plot a continuous line joining the discrete points. a), b) and c) {\bf As compression decreases, the variance of the signal noise increases or the correlation time of the noise decreases, the width and relative height of the exponential part of the kernel increases.} The optimal kernels consist of a \(\delta\) and an exponentially decaying curve. When \(\gamma\) is high (low \(\Ipred\) and \(\Ipast\) regime) or \(\sigma_\eta^2=0\), the exponential part disappears and the optimal kernel is a \(\delta\) function. Equally, when \(\tau_s\) is sufficiently low, the \(\delta\) function part diminishes, and the kernel becomes a decaying exponential. a) \(\sigma_\eta^2=2\) and \(\gamma=0.15\), in b) \(\tau_\eta=0.02\) and \(\gamma=0.15\) and in c) \(\tau_\eta=0.02\)s and \(\sigma_\eta^2=2\). In all \(\sigma_s^2=1\), \(\tau_s=2\)s, \(\sigma_\xi^2=1\), \(\tau=1\)s, \(dt=0.01\)s and \(T=1\)s. For all kernels except \(\sigma^2_\eta=0\), \(A(0)>10\) and compression noise is negligible. For \(\sigma_\eta^2=0\), \(A(0)\approx0.23\).}
    \label{KernelsAlone}
\end{figure}

\begin{figure}
    \centering
    \includegraphics[scale=0.4]{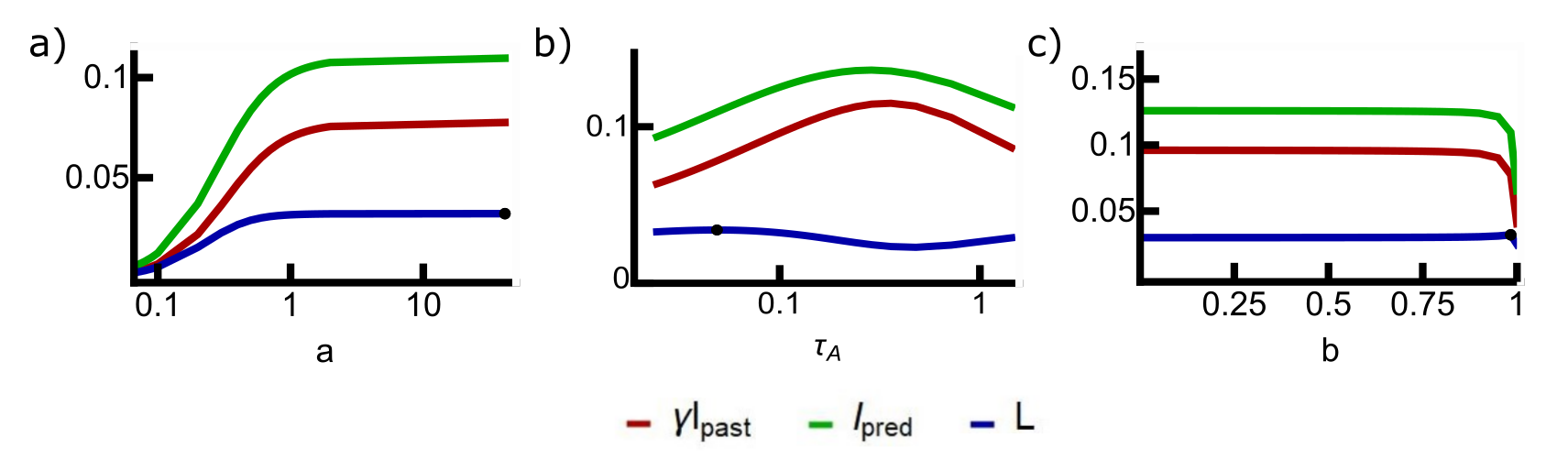}
    \caption{{\bf Dependence of predictive and past information on the integration kernel.} The form of the optimal kernel is: \(A_{\rm opt}(-t)=a_{\rm opt}\left(b_{\rm opt}\delta(t)+\frac{(1-b_{\rm opt})}{\tau_A^{\rm opt}} e^{-t/\tau_A^{\rm opt}}\right)\). Here, we calculate \(\Ipred\) and \(\Ipast\) for non-optimal kernels \(A(-t)=a\left(b\delta(t)+\frac{(1-b)}{\tau_A} e^{-t/\tau_{A}}\right)\) by varying one parameter at the time, the amplitude \(a\), the integration time of the kernel \(\tau_A\), or the relative height of the exponential part of the kernel \(b\), fixing the other parameters at their optimal values \(a_{\rm opt}=41.23\), \(b_{\rm opt}=0.984\) and \(\tau_A^{\rm opt}=0.049\)s. The optimal integration kernel corresponds to the black dots on the blue lines and to the coloured dots in figs. \ref{infcurves}a and b. In all graphs $\gamma = 0.2$. a) {\bf As the amplitude \(a\) increases, both \(\Ipred\) and \(\Ipast\) increase sharply to a plateau.} Recall the form of the compression process, \(x=\vec{\bf{A}}(\vec{\bf{s}}+\vec{\bf{\eta}})+\xi\). When the amplitude of the kernel \(a\) is small, the compression process is dominated by compression noise \(\xi\). As the amplitude increases, the effect of \(\xi\) is reduced and \(\Ipred\) and \(\Ipast\) increase. After the effect of compression noise \(\xi\) is diminished, the signal noise dominates. Because increasing \(a\) amplifies both the signal and the signal noise, it cannot diminish the effect of signal noise and increasing \(a\) further cannot increase \(\Ipred\) and \(\Ipast\). In this regime, the optimal amplitude is sufficiently high that compression noise \(\xi\) is negligible. The optimisation of the objective function \(\mathcal{L}\) therefore returns an arbitrary value greater than the point where \(\Ipred\) and \(\Ipast\) plateau, \(a\gtrsim 1\). b) {\bf As the integration time \(\tau_A\) increases, \(\Ipast\) and \(\Ipred\) increase to a maximum.} To mitigate signal noise, a system must time average, taking a weighted average over previous points on the trajectory. The integration time of the kernel \(\tau_A\) defines how much a system time averages. When \(\tau_A\) is zero, the system takes an instantaneous reading of the current value of the signal corrupted by signal noise. Here both \(\Ipred\) and \(\Ipast\) are low. As \(\tau_A\) increases, the effect of signal noise is reduced and \(\Ipred\) and \(\Ipast\) both increase. As \(\tau_A\) increases further, the dynamical error increases and reduces the ability of the system to predict. \(\Ipred\) and \(\Ipast\) therefore peak. c) {\bf Both \(\Ipred\) and \(\Ipast\) decrease with \(b\) } Both \(\Ipred\) and \(\Ipast\) decrease monotonically with the relative weighting of the \(\delta\) function \(b\), initially very slowly and then sharply as \(b\rightarrow 1\). Decreasing \(b\) decreases \(\Ipast\) marginally slower than \(\Ipred\), so the system chooses a high \(b_{\rm opt}\), prioritising the \(\delta\) peak over the exponential part of the kernel. In all panels, \(\sigma_\eta^2=2\), \(\sigma_s^2=1\), \(\sigma_\xi^2=1\), \(\tau_s=2\)s, \(\tau_\eta=0.02\)s, \(\tau=1\)s, \(dt=0.02\)s, \(T=2\)s.}
    \label{Optimum}
\end{figure}

\begin{figure}
    \centering
    \includegraphics[scale=0.4]{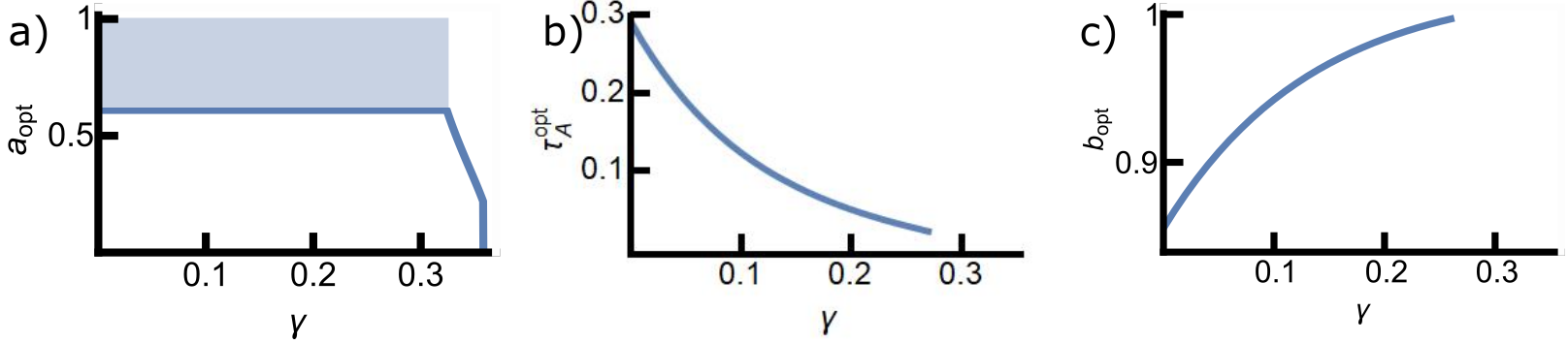}
    \caption{{\bf The parameters of the optimal integration kernel as the system moves along the information bound.}
    The optimal kernel is \(A_{\rm opt}(-t)=a_{\rm opt}\left(b_{\rm opt}\delta(t)+\frac{(1-b_{\rm opt})}{\tau_A^{\rm opt}} e^{-t/\tau_A^{\rm opt}}\right)\). In these graphs, we plot the optimal amplitude \(a\), integration time \(\tau_A\) and relative height of the exponential part of the kernel \(b\) against \(\gamma\). \(\gamma\) defines the level of compression. Low \(\gamma\) corresponds to the high \(\Ipred\) and \(\Ipast\) regime and high \(\gamma\) corresponds to the low \(\Ipred\) and \(\Ipast\) region. a) {\bf Amplitude quickly increases as \(\gamma\) decreases up to a plateau at \(\gamma\approx 0.32\).} At high compression \(\gamma\), the optimal amplitude is small, and \(\Ipast\) and \(\Ipred\) are low because the signal is dominated by the compression noise \(\xi\). As the compression level \(\gamma\) decreases, the system increases \(\Ipast\) and \(\Ipred\) by increasing the amplitude. However, beyond compression level \(\gamma\approx 0.32\), reducing the amplitude has little effect on \(\Ipred\) and \(\Ipast\) (see fig. \ref{Optimum}a), and the optimal amplitude increases so slowly that the numerical method can no longer find the optimum, instead giving a value which is bounded from below. Concomitantly, the amplitude is only well defined for compression level \(\gamma\gtrsim 0.32\). b) {\bf The integration time of the kernel increases as the compression level \(\gamma\) decreases.} At high compression level \(\gamma\), \(\tau_A^{\rm opt}\) is low, and the system does not time integrate, leading to low \(\Ipast\) and \(\Ipred\). As the compression level \(\gamma\) decreases, \(\tau_A^{\rm opt}\) increases, the system time averages more with greater resources.  c) {\bf The relative height of the \(\delta\) peak increases as \(\gamma\) decreases.} Inspecting fig. \ref{Optimum}c, we observe that as \(b\) increases, \(\Ipast\) decreases marginally faster than \(\Ipred\) at all \(\gamma\). This means that as \(\gamma\) increases and the compressive effect of \(\Ipast\) increases, the optimal \(b\) decreases towards zero. In all three graphs, the values of \(\tau_A\), \(a\) and \(b\) at zero compression \(\gamma=0\) correspond to their values in the Wiener filter. The amplitude of the Wiener kernel is therefore bounded from below rather than exactly defined. At sufficiently high compression \(\gamma\gtrsim 0.26\), the value of \(b\) becomes sufficiently close to one that the kernel becomes a \(\delta\) function and \(\tau_A\) becomes unstable. Similarly in the high \(\gamma\) region, the value of \(\tau_A\) becomes sufficiently small that \(e^{\frac{dt}{\tau_A}}\) becomes smaller than machine precision. In this case, the kernel is identical to a \(\delta\) function, and since amplitude is unbounded from above, \(b\) becomes unstable. There is, therefore, a region \(0.26\leq\gamma\leq0.32\) where all three quantities are poorly defined. In a), b) and c) \(\sigma_s^2=1\), \(\sigma_\xi^2=1\), \(\tau_s=2\)s, \(\tau_\eta=0.02\)s, \(\tau=1\)s, \(dt=0.02\)s, \(T=2\)s.}
    \label{changegamma}
\end{figure}

\begin{figure}
    \centering
    \includegraphics[scale=0.4]{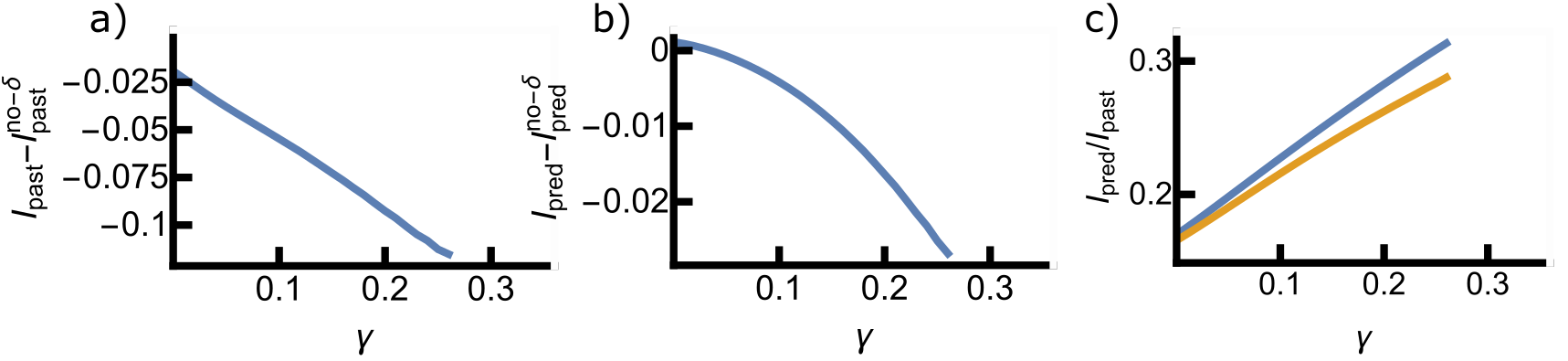}
    \caption{In these graphs, we compare the optimal kernel to an identical kernel omitting the \(\delta\) peak but leaving the rest of the kernel unchanged. a) {\bf Compared to the equivalent kernel with no \(\delta\) peak, the optimal kernel acquires less information about the past.} Recall that as the relative height of the \(\delta\) peak, \(b\), increases, the information collected about the past increases monotonically (fig. \ref{Optimum}d). Therefore, removing the \(\delta\) function always increases the information acquired about the past. b) {\bf Compared to the equivalent kernel with no \(\delta\) peak, the optimal kernel acquires less information about the future when the system is compressed (\(\gamma\) is large).} Only in the near uncompressed limit does adding a \(\delta\) peak increase \(\Ipred\). It is in this limit that the kernel is time averaging the most, so the \(\delta\) function is required to reduce dynamical error. As compression increases, the kernel time averages less, so the dynamical error goes down. This means adding the \(\delta\) no longer increases \(\Ipred\). c) {\bf The ratio \(\frac{\Ipred}{\Ipast}\) is always higher for the system with a \(\delta\) peak}. Regardless of whether \(\Ipred\) increases or decreases with the addition of a \(\delta\) peak, the ratio \(\frac{\Ipred}{\Ipast}\) always increases. Thus the information bottleneck always selects a kernel with a \(\delta\) peak.}
    \label{Nodelta}
\end{figure}

How does the optimal kernel compress the trajectory of the signal and the noise into a prediction of the future? How much information about the past and the future are retained in the compressed representation \(x\)? In this section, we examine the forms of optimal kernels \(\vec{\bf{A}}\) for discrete input signals with Markovian statistics. Once we have found the optimal kernels, we calculate the corresponding predictive information \(\Ipred\) and past information \(\Ipast\). Because these kernels are optimal, they will maximise the ratio of the predictive to the past information, \(\Ipred/\Ipast\), for a given system. For illustrative purposes, we will also calculate the predictive information \(\Ipred\) and past information \(\Ipast\) for an arbitrary kernel \(\vec{\bf{A}}\), but \(\Ipred/\Ipast\) will be lower for such non-optimal kernels. 

For a given signal and signal noise, there is an absolute limit on the amount of predictive information that can be extracted from a given amount of past information. Figure \ref{infcurves}a is a parametric plot of \(\Ipred\) and \(\Ipast\) for varying values of our compression variable \(\gamma\) (see Eq. \ref{Lagrangian2}). Here, each curve is the fundamental bound on \(\Ipred\) for a given \(\Ipast\), and for a given set of signals statistics: \(\sigma_s^2\), \(\tau_s\), \(\sigma_\eta^2\) and \(\tau_\eta\). At the top right of this information bound, the compression level \(\gamma\) is zero, and the system has maximum \(\Ipred\) and \(\Ipast\). Moving down the information bound, the system is compressed, and the system has access to reduced \(\Ipred\) and \(\Ipast\). Optimal kernels will result in values of \(\Ipred\) and \(\Ipast\) on the information bound, while arbitrary non-optimal kernels acting on signals with the same statistics will result in values of \(\Ipred\) and \(\Ipast\) below these bounds. The curves for \(\sigma_\eta^2=0\) have been calculated before \cite{chechik2003information}, but the other limits are new. We see that increasing the variance of the noise \(\sigma_\eta^2\) reduces the amount of predictive information \(\Ipred\) and past information \(\Ipast\) a system can extract at a given \(\gamma\) (fig. \ref{infcurves}a). At high \(\Ipred\) and \(\Ipast\), the predictive information extracted from a given amount of past information is indeed significantly lower when \(\sigma_\eta^2\) is higher. Moreover, the maximum \(\Ipast\) also decreases as \(\sigma_\eta^2\) becomes larger. Perhaps surprisingly, for lower \(\Ipred\) and \(\Ipast\), the predictive information that can be extracted for a given amount of past information is nearly independent of \(\sigma_\eta^2\).  Similarly, for \(\sigma_\eta^2>0\), increasing the correlation time of the noise also reduces the amount of predictive information \(\Ipred\) and past information \(\Ipast\) the system can extract from the signal trajectory. The ratio \(\Ipred/\Ipast\) is once again reduced with increased correlation time for high \(\Ipast\) while the ratio is constant at low \(\Ipast\) (fig. \ref{infcurves}b).

Examining the optimal kernel provides information about which characteristics of the signal are most important for predicting the future of the signal. As shown in fig. \ref{KernelsAlone}a, b and c, the optimal kernels take the form of a \(\delta\) function at \(t=0\) with a decaying exponential for \(t<0\):
\begin{equation}
    A_{\rm opt}(-t)=a_{\rm opt}\left(b_{\rm opt}\delta(t)+\frac{(1-b_{\rm opt})}{\tau_A^{\rm opt}} e^{-t/\tau_A^{\rm opt}}\right).
\end{equation}
The \(\delta\) peak prioritises the signal's current value, but the exponential function averages over the trajectory, with lower weight given to time points further back into the past.  Here \(a_{\rm opt}\) is the amplitude of the entire kernel, \(b_{\rm opt}\) is the weighting of the \(\delta\) peak relative to the exponential part of the kernel, and \(\tau_{A}^{\rm opt}\) is the decay rate of the kernel. We normalise the exponential part of the kernel with the decay rate. We note that while we write down a continuous form of the kernel, the kernel itself is discrete and only defined at integer multiples of the timestep \(dt\).

The shape of the kernel changes along the information bound. Consider the system where \(\sigma_\eta^2=2\) and \(\tau_\eta=0.02\)s (fig. \ref{infcurves}a, middle blue line, fig. \ref{infcurves}b, middle red line). At the top right of the information bound, the compression level \(\gamma=0\), the system is uncompressed and can access maximum \(\Ipred\) and \(\Ipast\). In this limit, the kernel is a slowly decaying exponential supplemented by a \(\delta\) function (fig.\ref{KernelsAlone}c, dashed line). Initially, as we move down the information bound, increasing $\gamma$, the width and relative height of the exponential function decrease until the function becomes a \(\delta\) function (fig.\ref{KernelsAlone}c, light yellow line). Only then does the amplitude of the whole kernel decrease to zero.

To understand the optimal shape of the kernel, we need to understand the origins of the fluctuations in the output because these fluctuations limit the accuracy of prediction. Two of these we have already discussed: signal noise and compression noise, modelled by $\eta$ and $\xi$ in Eq. \ref{output}, respectively. Signal noise causes errors in the signal at the point of detection. Compression noise corrupts the output of the compression process. The final source of fluctuations in the output is known as the ``dynamical error''~\cite{malaguti2021theory}. It arises from time integration. Due to time integration, the output depends on input values further back into the past, which are less correlated with the current input \cite{malaguti2021theory}.


To understand how a system can mitigate these sources of error, we note that the compressed output is given by \(x=\vec{\bf A}(\vec{\bf s}+\vec{\bf \eta})+\xi\). Increasing the amplitude of the kernel can mitigate the effect of the compression noise \(\xi\) by amplifying the signal over the compression noise. Changing the amplitude cannot reduce the effect of the signal noise \(\vec{\bf \eta}\) because the signal and noise will be amplified together. Signal noise must be mitigated by time averaging. By using more independent time points further into the past, the system can better estimate the current value of the signal. The integration time \(\tau_{A}\) sets the width of the kernel and the window over which time averaging is performed. However, using time points further back into the past introduces dynamical error, which is mitigated by prioritising more recent values over values further into the past. Mitigating signal noise and dynamical error thus put opposing requirements on the integration time, leading to an optimum in \(\tau_A\) \cite{becker2015optimal,malaguti2021theory}.

We next ask how varying the key parameters of the kernel: \(a\), \(b\) and \(\tau_A\), affects these error sources. Answering this question will clarify how these parameters affect the past and predictive information \(\Ipred\) and \(\Ipast\), which in turn helps us understand how the optimal kernel's shape varies along the information bounds shown in fig. \ref{infcurves}a and b. We generate a set of non-optimal kernels for a given signal, \(A(t)=a\left(b\delta(t)+\frac{(1-b)}{\tau_A} e^{-\frac{1}{\tau_A}t}\right)\), by varying \(a\), \(b\) and \(\tau_A\) away from the optimum. To separate the effects of varying these three quantities, we fix two of the three quantities \(a=a_{\rm opt}\), \(b=b_{\rm opt}\) and \(\tau_A=\tau_A^{\rm opt}\), and vary the other.

How does varying the amplitude, \(a\), allow the system to mitigate our three error types? Recall the expression for the compressed output: \(x=\vec{\bf{A}}(\vec{\bf{s}}+\vec{\bf{\eta}})+\xi\). When the amplitude of \(\vec{\bf{A}}\), \(a\), is small, the compression noise \(\xi\) dominates the signal. In this case, both \(\Ipred\) and \(\Ipast\) are small (fig. \ref{Optimum}a). As \(a\) increases, both \(\Ipred\) and \(\Ipast\) increase as the kernel amplifies the corrupted signal over the compression noise. Eventually, the compression noise becomes negligible compared to the propagated input noise, and \(\Ipred\) and \(\Ipast\) plateau as a function of \(a\). Indeed, while changing \(a\) can lift the signal above the compression noise \(\xi\), it cannot mitigate the effect of signal noise \(\eta\) because the kernel amplifies the signal and input noise together. Similarly, as increasing the amplitude of the kernel does not affect how different points in the trajectory are weighted relatively in the kernel, it cannot decrease dynamical error.

Since varying the amplitude cannot mitigate signal noise, it can only be mitigated by varying the relative height and width of the exponential part of the kernel. The exponential part of the kernel takes a non-uniform time average over those time points in the past, mitigating signal noise. Consider first the integration time of the kernel, set by \(\tau_{A}\). As \(\tau_A\) increases and the kernel widens, both \(\Ipred\) and \(\Ipast\) initially increase as the system averages out the signal noise (fig. \ref{Optimum}c). They then peak at two different optimal integration times, which arise from the trade-off between minimizing the dynamical error and time averaging\cite{malaguti2021theory}. 

We next ask how \(\Ipred\) and \(\Ipast\) change with the relative importance of the \(\delta\) peak: \(b\). Initially, \(\Ipred\) and \(\Ipast\) decrease very slowly as the relative importance of the \(\delta\) peak increases. As \(b\) approaches one, both quantities drop sharply. For all values of the compression level \(\gamma\), having a \(\delta\) peak decreases the amount of past information the system obtains with the kernel (fig. \ref{Nodelta}a). For all but the lowest values of the compression level \(\gamma\), having a \(\delta\) peak also decreases the amount of predictive information the system obtains with the kernel (fig. \ref{Nodelta}b). Only in the zero compression limit does adding a \(\delta\) peak increase \(\Ipred\); in the SI, we prove that this is true even as \(dt\rightarrow 0\).  In this limit, the system finds the optimal trade-off between minimising signal noise via a wide integration kernel and minimising dynamical error via a \(\delta\) peak (the compression noise is negligible). The \(\delta\) peak emphasises the most recent signal value, the signal value most correlated with the future point the system is trying to predict. In this limit, \(\Ipred\) peaks at \(b=b_{\rm wiener}\) (fig. \ref{Nodelta}d, dashed lines). 

Since both \(\Ipast\) and \(\Ipred\) (except for the uncompressed limit) decrease upon adding a \(\delta\) peak, a pertinent question rises is why the optimal kernel of the system at the information bound features a \(\delta\) peak at all. The answer is that \(\Ipast\) decreases more than \(\Ipred\) upon adding a \(\delta\) peak, so that the ratio \(\Ipred / \Ipast\) increases. This effect is strongest in the compressed regime (fig. 5c), which explains why the \(\delta\) peak is most pronounced in the high \(\gamma\) regime of strong compression.   

We can now understand the shape of the kernel along the information bottleneck curves (fig. \ref{changegamma}). We start in the highly compressed region where 
\(\Ipast\) and \(\Ipred\) are low, because the amplitude of the kernel $a$ is low (Fig \ref{changegamma}a) and the compression noise is relatively large. Because of the latter, the effect of the signal noise is relatively small. This means that time averaging is not important. The optimal integration time will be short because that minimizes the dynamical error (Fig \ref{changegamma}c). The $\delta$ term will be relatively large (fig. \ref{changegamma}b) because increasing the \(\delta\) peak maximises the objective function by decreasing \(\Ipast\) more than \(\Ipast\).  

To increase \(\Ipast\) and \(\Ipred\) (corresponding to decreasing $\gamma$), the amplitude of the kernel must rise so that signal is lifted above the compression noise (fig. \ref{changegamma}). Because the kernel acts on both the signal and the signalling noise but not the compression noise, this inevitably makes the effect of the signal noise stronger than the compression noise. This means that time averaging becomes more important, which in turn necessitates a longer integration time (fig. \ref{changegamma}c). Since increasing \(\tau_A\) also increases the magnitude of the kernel, amplifying the signal and signalling noise over the compression noise, the relative importance of the $\delta$-peak contribution falls.  

In the regime of high \(\Ipred\) and \(\Ipast\) (low $\gamma$), the compression noise has become negligible, and the output noise is caused by a combination of signal noise and dynamical error. 
The optimal integration time in the uncompressed limit arises from the trade-off between the two error types. Similarly, since adding a \(\delta\) peak reduces dynamical error, this trade-off also sets the optimal relative height of the exponential part of the kernel and the \(\delta\) peak.  
Hence the numerical procedure no longer finds a unique solution for the amplitude, it only ensures that it is large enough. In the limit $\gamma \to 0$, the kernel becomes identical to that given by the Wiener filter, as we show in Appendix \ref{Wiener}. The Wiener filter has been used to analyse optimal kernels for Markovian signals~\cite{becker2015optimal}, although that study did not address the effect of correlations in the noise.

\begin{figure}
    \centering
    \includegraphics[scale=0.4]{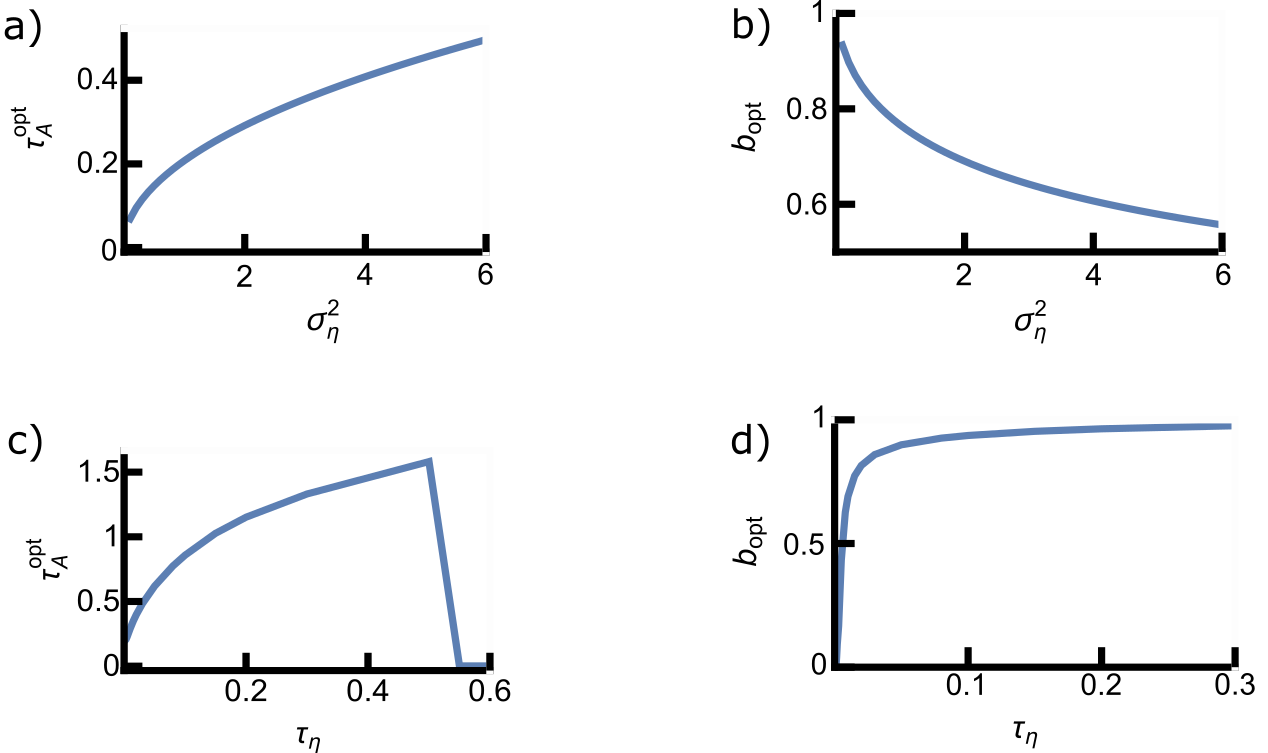}
    \caption{a) {\bf The system integrates over a longer time interval as the noise variance increases.} The width of the kernel at compression level \(\gamma=0\) (max \(\Ipred\) and \(\Ipast\)) is a tradeoff between minimising the effect of signalling noise and minimising dynamical error. Widening the kernel increases dynamical error but decreases the effect of signalling noise. As \(\sigma^2_\eta\) increases, the effect of signalling noise increases relative to dynamical error, and so the optimal kernel will be wider. b) {\bf The system prioritises time averaging over instantaneous measurement as the noise variance increases.} The tradeoff between minimising signalling noise and dynamical error also sets the optimal ratio between the exponential part of the kernel and the \(\delta\) peak. Decreasing the height of the \(\delta\) peak by decreasing \(b\) decreases the effect of signalling noise and so is prioritised as the noise variance increases. c) {\bf The system integrates over a longer time interval as the correlation time of the noise increases.} Correlated noise requires a wider window of time averaging relative to uncorrelated noise because of the persistence of the correlations. Thus the integration time of the kernel increases as the correlation time of the noise increases. Yet, when the integration time becomes too long compared to the input correlation time, time integration not only averages out the noise in the signal, but also the signal itself. Beyond this point, the system gives up the mechanism of time integration and instead becomes an instantaneous responder: \(\tau_A^{\rm opt}\) rapidly drops to zero. d) {\bf The system prioritises an instantaneous measurement over time averaging as the correlation time of the noise increases.} Since the kernel widens as the correlation time of the noise increases, the dynamical error increases. Thus the system prioritises the \(\delta\) function by decreasing \(b\) as \(\tau_\eta\) increases. For all graphs \(\gamma=0\), \(\sigma^2_\xi= 1\), \(\sigma^2_s=1\), \(\tau_s=2\)s, \(\tau=1\)s, \(dt=0.02\)s and \(T=2\)s.}
    \label{TauAndSigma}
\end{figure}

Now that we understand the optimal shape of the integration kernel, we are in a position to understand the effects of varying the magnitude and the correlation time of the input noise, \(\sigma^2_\eta\) and \(\tau_\eta\), respectively. The correlation time of the exponential part of the kernel increases and the relative weight of the \(\delta\) peak decreases with \(\sigma_{\eta}^2\) because more signal noise requires more time averaging (fig. \ref{TauAndSigma}a and b). In the absence of noise, \(\sigma_{\eta}^2=0\), the kernel takes the form of a \(\delta\) function because, for a Markovian signal, all the predictive power is stored in the current signal value. Indeed, in all of our systems, time averaging is performed to better estimate the current signal, which is then maximally predictive of the future signal.

The kernel also changes with the noise's correlation time. To mitigate the effects of correlated noise, the system must time average over periods longer than the correlation time of the noise, but shorter than the correlation time of the signal: \(\tau_s>\tau_A>\tau_\eta\). Initially, as \(\tau_\eta\) increases, the width of the exponential part of the kernel \(\tau_A\) increases  (fig. \ref{TauAndSigma}c). As \(\tau_\eta\rightarrow\tau_s\), the width of the kernel decreases because the system can no longer average out the noise without averaging out the signal. This also explains why the relative importance of the exponential filter decreases and that of the \(\delta\) peak increases as \(\tau_\eta\rightarrow\tau_s\) (fig. \ref{TauAndSigma}d). Conversely, decreasing the input correlation time prioritises the exponential filter. Indeed, Becker {\it et al.} derived using Wiener filtering theory the optimal integration function for signals with $\delta$ correlated input noise, corresponding to $\tau_\eta \to 0$, and found that the optimal kernel is a simple exponential filter \cite{becker2015optimal}. Lastly, we note that when the correlation time of the signal noise becomes comparable to the correlation time of the signal itself,  \(\tau_\eta \sim \tau_s\), the system cannot time average out the signal noise without time averaging out the signal itself. The system cannot do better than taking an instantaneous kernel, and the relative height of the exponential part of the kernel goes to zero (fig. \ref{TauAndSigma}d). This behaviour reflects that observed for cellular signaling systems \cite{malaguti2021theory}.

\section{The push-pull network}\label{pushpullsect}

\begin{figure}
    \centering
    \includegraphics[scale=0.4]{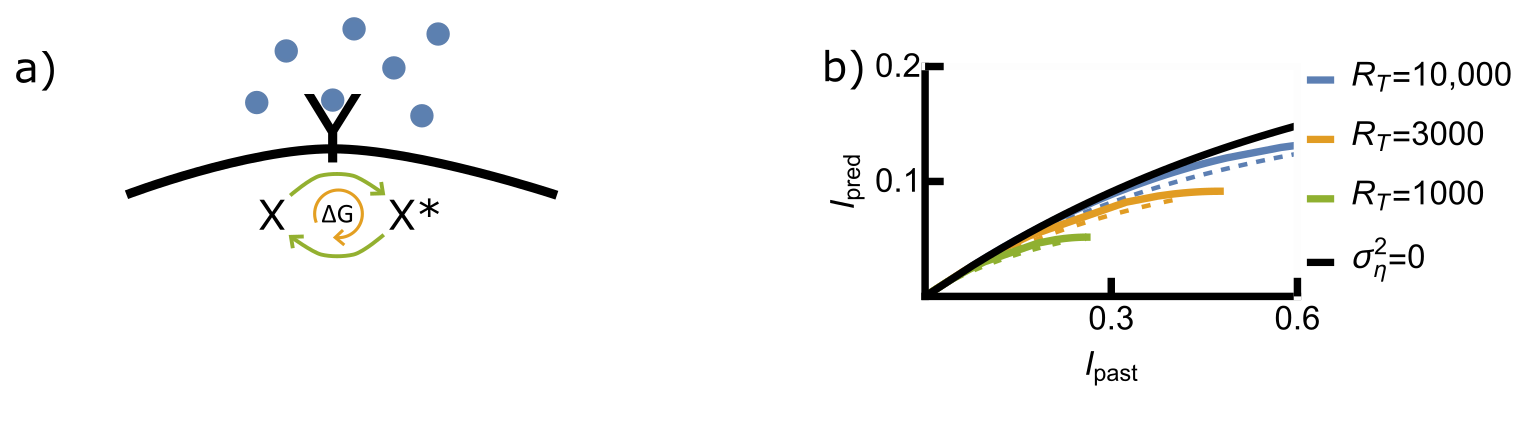}
    \caption{a) {\bf A summary of the push-pull motif.} Receptors on the outside of a cell interact with ligand molecules in the environment. ligand molecules bind and unbind to receptors. Inside the cell, output molecules diffuse in and out of contact with the receptors. Output molecules in contact with ligand-bound receptors become activated. Output molecules in the solution spontaneously deactivate. There are a total number \(X_T\) output molecules. b) {\bf The push-pull motif lies below the information curves for all receptor number values.} Using equation \ref{sigmaeta}, we can calculate optimal information bounds (solid lines) at constant noise, which can be directly compared to the curves generated using the push-pull kernel (dashed lines). We see that the information curves traced using the push-pull kernels lie below the information curves for all \(\sigma_\eta^2\propto\frac{1}{R_T}\) and \(\sigma_s^2=2\times10^{-4}\).}
    \label{pushpullcompare}
\end{figure}

\begin{figure}
    \centering
    \includegraphics[scale=0.4]{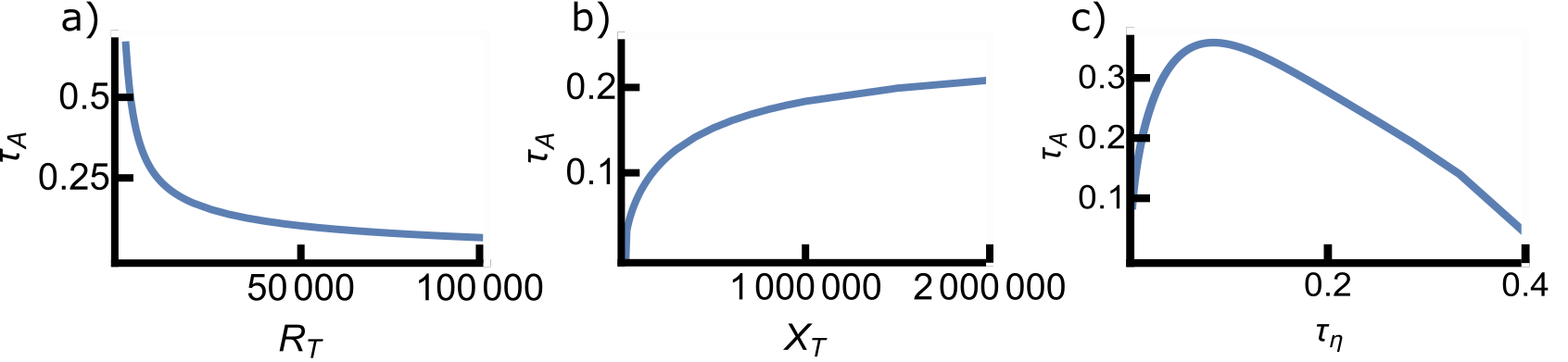}
    \caption{a) {\bf The push-pull network has a wider kernel for larger noise variance (smaller \(R_T\))}. We plot the correlation time of the kernels found for the push-pull network using a discretised version of the method from~\cite{tjalma} (SI) for varying \(R_T\) (corresponding to varying \(\sigma_{\eta}^2\)). We see that as \(R_T\) increases and signal noise variance decreases, the width of the optimal kernel decreases. Thus the push-pull kernel uses time averaging to suppress signal noise, like the optimal kernel. b) {\bf The push-pull network has a narrower kernel for greater compression (smaller \(X_T)\).} We repeat the method increasing \(X_T\) which is comparable to decreasing the compression level \(\gamma\). Here \(\tau_\eta=0.02\) and \(R_{T}=100\) corresponds to \(\sigma_{\eta}^2=0.032\). The correlation time of the optimal kernels decreases as resources decrease and compression increases. Thus noise suppression through time integration is resource intensive for the push-pull network. c) {\bf The push-pull kernel initially widens and then narrows as noise correlation time increases (\(\tau_\eta\))}. We repeat the method increasing \(\tau_\eta\). We see that as \(\tau_\eta\) increases, the width of the optimal kernel initially increases and then decreases. In all graphs the average proportion of ligand-bound receptors and activated output molecules is \(\phi_l=\phi_x=0.5\) respectively, and the average concentration of ligands \(c=1\). In all, \(\sigma_s^2=2\times10^{-4}\). In a) and b) \(\tau_\eta=0.02\)s. In b) and c) \(R_T=1\times10^4\) and in a) and c) \(X_T=1\times10^{12}\).}
    \label{PushPullKernel}
\end{figure}

Having calculated the properties of optimal kernels, we now wish to compare our results to a standard signal-processing motif in biology: the push-pull motif. The cell must detect and predict the concentration of ligand molecules in the environment. The push-pull motif consists of receptors on the surface of a cell that detect the concentration of ligand molecules in the environment by binding to them (fig.\ref{pushpullcompare}a). Inside the cell, output molecules diffuse in and out of contact with the receptors. Output molecules in contact with bound receptors are activated, using ATP to drive the reaction. These molecules then spontaneously deactivate over time. The number of activated output molecules reflects the number of bound receptors, allowing the cell to estimate the concentration of ligand molecules outside the cell. Intrinsic to the push-pull network is correlated signal noise caused by the binding of ligand molecules to receptors. We are interested in whether the push-pull kernel can mitigate this noise.

The push-pull kernel is an exponential function: \(A(-t)\propto e^{-t/\tau_A}\)~\cite{tjalma}. Here, \(\tau_A\) is the integration time of the kernel. In the supplementary information, we extract the variance of the signal noise from the push-pull system (SI, equation \ref{sigmaeta}), which we simplify to:
\begin{align}
     \sigma_{\eta}^2\propto\frac{1}{R_T}
     \label{noiseinpushpull}
\end{align}
to aid understanding.
Here \(R_T\) is the total number of receptors. Additionally, for the push-pull motif, resources, and therefore compression level, are dictated by the number of receptors \(R_T\) and the number of output molecules \(X_T\). 

In what follows, we increase the compression level by reducing \(X_T\) while keeping \(R_T\) constant to keep the signal noise variance constant. Fig. \ref{pushpullcompare}b shows that the information curves traced by the push-pull kernel (dashed lines) fall below those of the optimal kernels (solid line). However, the difference is small, hinting that the push-pull network is nearly optimal. To analyze this further, we study the integration kernels. 

We find that, just like the optimal kernels, the optimal kernels for the push-pull motif widen with the variance of the signal noise and narrow with compression. Fig. \ref{PushPullKernel}a shows that when \(R_T\) decreases, which increases the signal noise variance, the push-pull kernels widen. In contrast, when \(X_T\) decreases, which increases the compression level, the kernels narrow (fig. \ref{PushPullKernel}b). Thus, the push-pull kernel uses time averaging to mitigate signal noise, and the ability to time average is reduced by compression, like the optimal kernels (fig. \ref{changegamma}).

 In ref. ~\cite{malaguti2021theory}, the authors observe that cells using the push-pull motif can reduce the sensing error by either increasing the number of receptors \(R_T\) or by taking more measurements per receptor via the mechanism of time integration (increasing \(\tau_A\)). These two statements can now be directly related to signal noise. Increasing the number of receptors reduces the signal noise \(\sigma_{\eta}^2\), while time integration corresponds to widening the kernel to average out signal noise. In the same study, the authors observe an optimal integration time which increases as the number of receptors decreases. Moreover, they found that the optimal integration time decreases for larger compression noise, i.e., smaller \(X_T\). Our results corroborate and explain these findings (see figs. \ref{TauAndSigma}a and \ref{PushPullKernel}a).

Additionally, in ref. ~\cite{malaguti2021theory}, the authors observe that cells using the push-pull motif respond to the correlation time of the noise increasing by initially increasing the integration time of the kernel (fig. \ref{PushPullKernel}c). Then as \(\tau_\eta\) approaches \(\tau_s\), time integration averages out the signal as well as the noise, lowering the utility of time integration as a strategy. As \(\tau_\eta\) increases further, the optimal integration time gradually decreases back to zero. Here, the push-pull network's best strategy is merely to capture the current value of the signal, despite noise corruption. In the limit that \(X_T\) is so large as to be effectively infinite, rendering the compression noise negligible, the integration time decreases slowly beyond the peak. For a smaller \(X_T\) where compression noise is finite, this drop is sharp (SI, fig. \ref{SItau}), mirroring the equivalent drop found for our optimal system (fig. \ref{TauAndSigma}d).

The push-pull kernel does not and cannot manifest a \(\delta\) peak. The system instead has an exponentially decaying kernel alone. As \(\tau_\eta\rightarrow 0\), the theoretical optimal kernels also tend towards an exponentially decaying kernel without a \(\delta\) peak (fig. \ref{KernelsAlone}a). We suggest, therefore, that the push-pull kernel is optimised for signal noise with a short correlation time. Nonetheless, while the system cannot replicate the \(\delta\) peak, the push-pull kernel still attempts to mitigate correlated signal noise in other ways. Specifically, the kernel widens as \(\tau_\eta\) increases (fig. \ref{PushPullKernel}c), as observed for the optimal kernels (fig. \ref{TauAndSigma}c).

What does the push-pull system lose by not being able to implement a \(\delta\) peak? For all but the lowest values of the compression level \(\gamma\), the kernel without a \(\delta\) peak collects more predictive and past information (fig. \ref{Nodelta}a and b) than a kernel with a \(\delta\) peak. The \(\delta\) peak emerges only when the predictive information is maximized under the specific constraint of limiting past information. As discussed in \cite{Tjalma.2023}, maximizing predictive information while constraining past information will  yield systems that differ from those that maximize predictive information under the constraint of resource cost in terms of protein copies and energy. It is conceivable that the latter would not yield a \(\delta\) peak. 


A biological system could hypothetically create a network capable of implementing a kernel much closer in shape to our theoretically optimal kernels. Creating an additional \(\delta\) peak would require coupling two push-pull motifs in parallel, with different turnover rates of the readout \cite{govern2012fundamental}. The faster push-pull motif would provide a sharp spike in the kernel close to \(t=0\)s, which would approximate a \(\delta\) function, while the slower one would act as the exponentially decaying part of the kernel. A motif such as this is resource intensive, so the limited potential advantages may explain why two such parallel push-pull networks have not yet been observed in cellular systems.

\section{Conclusions}

Time averaging is essential for accurately detecting and predicting the true values of signals corrupted by signal noise. An optimal system will vary the width of the kernel to compensate for different characteristics of this signal noise, widening it for a greater variance or longer correlation times and shortening it if the opposite is true. Where the noise characteristics demand it, most notably when the correlation time of the noise is long, kernels will widen to the extent that dynamical error becomes a concern for the system. In this case, an optimal system will add a \(\delta\) peak in the kernel at the current time. The push-pull kernel replicates the optimal kernel for systems where the noise correlation time is very short but otherwise fails as it cannot replicate a \(\delta\) peak.

Suppose a system has finite resources for prediction. In that case, its ability to time average is reduced—both the theoretically optimal kernels and those of the push-pull motif narrow as their resources are restricted at fixed signal noise. With sufficient compression, the optimal kernels will only collect the most recent time point, omitting time averaging completely. In such cases, the system cannot mitigate the effect of signal noise at all.

We have combined the information bottleneck and the Wiener filter to study these systems. This technique can be applied to more complex signals, such as those described by the generalised Langevin equation~\cite{sachdeva2021optimal}. Studying how noise corruption affects techniques for processing more complex signals is the subject of further work.

\bibliographystyle{unsrt}
\bibliography{Bibliography.bib}

\begin{thebibliography}{10}

\bibitem{sachdeva2021optimal}
Vedant Sachdeva, Thierry Mora, Aleksandra~M Walczak, and Stephanie~E Palmer.
\newblock Optimal prediction with resource constraints using the information
  bottleneck.
\newblock {\em PLoS computational biology}, 17(3):e1008743, 2021.

\bibitem{malaguti2021theory}
Giulia Malaguti and Pieter~Rein Ten~Wolde.
\newblock Theory for the optimal detection of time-varying signals in cellular
  sensing systems.
\newblock {\em Elife}, 10:e62574, 2021.

\bibitem{mitchell2009adaptive}
Amir Mitchell, Gal~H Romano, Bella Groisman, Avihu Yona, Erez Dekel, Martin
  Kupiec, Orna Dahan, and Yitzhak Pilpel.
\newblock Adaptive prediction of environmental changes by microorganisms.
\newblock {\em Nature}, 460(7252):220--224, 2009.

\bibitem{tagkopoulos2008predictive}
Ilias Tagkopoulos, Yir-Chung Liu, and Saeed Tavazoie.
\newblock Predictive behavior within microbial genetic networks.
\newblock {\em science}, 320(5881):1313--1317, 2008.

\bibitem{berg1977physics}
Howard~C Berg and Edward~M Purcell.
\newblock Physics of chemoreception.
\newblock {\em Biophysical journal}, 20(2):193--219, 1977.

\bibitem{Bialek2005}
William Bialek and Sima Setayeshgar.
\newblock {Physical limits to biochemical signaling.}
\newblock {\em Proc. Natl. Acad. Sci. U.S.A.}, 102(29):10040--10045, 2005.

\bibitem{Wang2007}
Kai Wang, Wouter~Jan Rappel, Rex Kerr, and Herbert Levine.
\newblock {Quantifying noise levels of intercellular signals}.
\newblock {\em Phys. Rev. E}, 75(6):061905, 2007.

\bibitem{Rappel2008}
Wouter~Jan Rappel and Herbert Levine.
\newblock {Receptor noise and directional sensing in eukaryotic chemotaxis}.
\newblock {\em Phys. Rev. Lett.}, 100(22):228101, 2008.

\bibitem{govern2012fundamental}
Christopher~C Govern and Pieter~Rein ten Wolde.
\newblock Fundamental limits on sensing chemical concentrations with linear
  biochemical networks.
\newblock {\em Physical review letters}, 109(21):218103, 2012.

\bibitem{Mehta2012}
Pankaj Mehta and David~J. Schwab.
\newblock {Energetic Costs of Cellular Computation}.
\newblock {\em Proc. Natl. Acad. Sci. U.S.A.}, 109(44):17978, 2012.

\bibitem{govern2014}
Christopher~C. Govern and Pieter~Rein ten Wolde.
\newblock {Optimal resource allocation in cellular sensing systems}.
\newblock {\em Proc. Natl. Acad. Sci. U.S.A.}, 111(49):17486--17491, 2014.

\bibitem{Govern2014b}
Christopher~C. Govern and Pieter~Rein {ten Wolde}.
\newblock {Energy Dissipation and Noise Correlations in Biochemical Sensing}.
\newblock {\em Phys. Rev. Lett.}, 113:258102, 2014.

\bibitem{Kaizu2014}
Kazunari Kaizu, Wiet {De Ronde}, Joris Paijmans, Koichi Takahashi, Filipe
  Tostevin, and Pieter Rein~Ten Wolde.
\newblock {The Berg-Purcell limit revisited}.
\newblock {\em Biophys. J.}, 106(4):976--985, 2014.

\bibitem{Fancher:2017ba}
Sean Fancher and Andrew Mugler.
\newblock {Fundamental Limits to Collective Concentration Sensing in Cell
  Populations}.
\newblock {\em Phys. Rev. Lett.}, 118(7):078101, February 2017.

\bibitem{Tostevin2009}
Filipe Tostevin and Pieter~Rein {ten Wolde}.
\newblock {Mutual information between input and output trajectories of
  biochemical networks}.
\newblock {\em Phys. Rev. Lett.}, 102:218101, 2009.

\bibitem{2021.Mattingly}
H.~H. Mattingly, K.~Kamino, B.~B. Machta, and T.~Emonet.
\newblock {Escherichia coli chemotaxis is information limited}.
\newblock {\em Nature Physics}, 17(12):1426--1431, 2021.

\bibitem{Hinczewski:2014iq}
Michael Hinczewski and D~Thirumalai.
\newblock {Cellular Signaling Networks Function as Generalized
  Wiener-Kolmogorov Filters to Suppress Noise}.
\newblock {\em Physical Review X}, 4(4):3--15, October 2014.

\bibitem{becker2015optimal}
Nils~B Becker, Andrew Mugler, and Pieter~Rein Ten~Wolde.
\newblock Optimal prediction by cellular signaling networks.
\newblock {\em Physical review letters}, 115(25):258103, 2015.

\bibitem{mora2010limits}
Thierry Mora and Ned~S Wingreen.
\newblock Limits of sensing temporal concentration changes by single cells.
\newblock {\em Physical review letters}, 104(24):248101, 2010.

\bibitem{Lang:2014ir}
Alex~H Lang, Charles~K Fisher, Thierry Mora, and Pankaj Mehta.
\newblock {Thermodynamics of Statistical Inference by Cells}.
\newblock {\em Phys. Rev. Lett.}, 113(14):148103, October 2014.

\bibitem{Mora:2019fd}
Thierry Mora and Ilya Nemenman.
\newblock {Physical Limit to Concentration Sensing in a Changing Environment}.
\newblock {\em Physical Review Letters}, 123(19):198101, November 2019.

\bibitem{wiener1949extrapolation}
Norbert Wiener, Norbert Wiener, Cyberneticist Mathematician, Norbert Wiener,
  Norbert Wiener, and Cybern{\'e}ticien Math{\'e}maticien.
\newblock {\em Extrapolation, interpolation, and smoothing of stationary time
  series: with engineering applications}, volume 113.
\newblock MIT press Cambridge, MA, 1949.

\bibitem{kolmogorov1962interpolation}
Andre{\u\i}~Nikolaevich Kolmogorov.
\newblock {\em Interpolation and extrapolation of stationary random sequences}.
\newblock Izv. Akad. Nauk SSSR., Ser. Mater., 1941.

\bibitem{ShannonWiener}
H.W. Bode and C.E. Shannon.
\newblock A simplified derivation of linear least square smoothing and
  prediction theory.
\newblock {\em Proceedings of the IRE}, 38(4):417--425, 1950.

\bibitem{Tishby1999}
Naftali Tishby, Fernando~C. Pereira, and William Bialek.
\newblock {The Information Bottleneck Method}.
\newblock In {\em 37th Annu. Allert. Conf. Commun. Control. Comput.}, pages
  368--377, 1999.

\bibitem{chechik2003information}
Gal Chechik, Amir Globerson, Naftali Tishby, and Yair Weiss.
\newblock Information bottleneck for gaussian variables.
\newblock {\em Advances in Neural Information Processing Systems}, 16, 2003.

\bibitem{bauer2021trading}
Marianne Bauer, Mariela~D Petkova, Thomas Gregor, Eric~F Wieschaus, and William
  Bialek.
\newblock Trading bits in the readout from a genetic network.
\newblock {\em Proceedings of the National Academy of Sciences},
  118(46):e2109011118, 2021.

\bibitem{bauer2022does}
Marianne Bauer.
\newblock How does an organism extract relevant information from transcription
  factor concentrations?
\newblock {\em Biochemical Society Transactions}, 50(5):1365--1376, 2022.

\bibitem{palmer2015predictive}
Stephanie~E Palmer, Olivier Marre, Michael~J Berry, and William Bialek.
\newblock Predictive information in a sensory population.
\newblock {\em Proceedings of the National Academy of Sciences},
  112(22):6908--6913, 2015.

\bibitem{Matz}
Michael Meidlinger, Andreas Winkelbauer, and Gerald Matz.
\newblock On the relation between the gaussian information bottleneck and
  mse-optimal rate-distortion quantization.
\newblock In {\em 2014 IEEE Workshop on Statistical Signal Processing (SSP)},
  pages 89--92, 2014.

\bibitem{Goldbeter1981}
A~Goldbeter and D~E Koshland.
\newblock {An amplified sensitivity arising from covalent modification in
  biological systems.}
\newblock {\em Proc. Natl. Acad. Sci. U. S. A.}, 78(11):6840--6844, 1981.

\bibitem{Alon:2007tz}
Uri Alon.
\newblock {\em Introduction to Systems Biology: Design Principles of Biological
  Networks}.
\newblock CRC press, Boca Raton, FL, 2007.

\bibitem{cryer2008time}
Jonathan~D Cryer and Kung-Sik Chan.
\newblock {\em Time series analysis: with applications in R}, volume~2.
\newblock Springer, 2008.

\bibitem{tjalma}
Age~J. Tjalma, Vahe Galstyan, Jeroen Goedhart, Lotte Slim, Nils~B. Becker, and
  Pieter Rein~ten Wolde.
\newblock Trade-offs between cost and information in cellular prediction, 2023.

\bibitem{Tjalma.2023}
Age~J Tjalma, Vahe Galstyan, Jeroen Goedhart, Lotte Slim, Nils~B Becker, and
  Pieter Rein~ten Wolde.
\newblock {Trade-offs between cost and information in cellular prediction}.
\newblock {\em arXiv}, 2023.

\end{thebibliography}

\beginsupplement

\section*{Supplementary Material}

\section{The kernel shape (but not amplitude) and information bottleneck curves are independent of compression noise}

\begin{figure}[h!]
     \centering
     \includegraphics[scale=0.4]{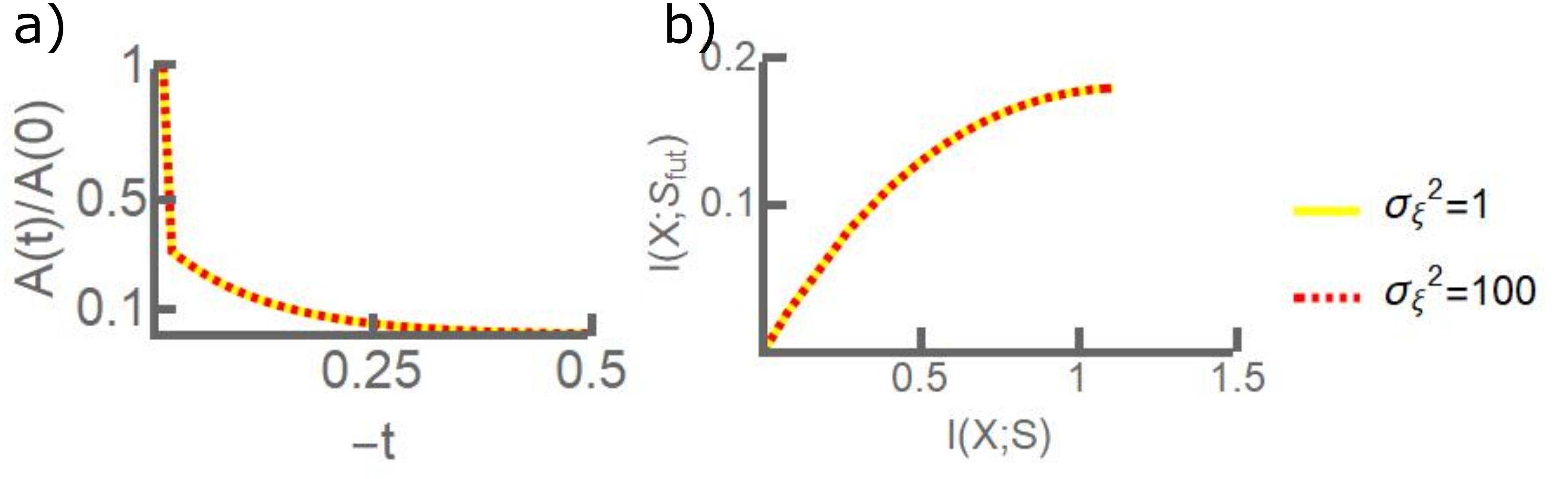}
     \caption{We compare the rescaled kernels (a) and information curves (b) for a system with \(\sigma_\xi^2=1\) and \(\sigma_\xi^2=100\) to see that they are identical. The kernels will be amplified for higher \(\sigma_\xi^2\), but the shape will not change. Here \(\sigma_{\eta}^2=2\), \(\tau_s=2\)s, \(\tau=1\)s, \(dt=0.01\)s, \(T=0.5\)s. For the kernels 
     \(\gamma=0.1\). }
     \label{XiCompare}
 \end{figure}

In figure \ref{XiCompare}, we compare the rescaled kernels (a) and information curves (b) for a system with \(\sigma_\xi^2=1\) and \(\sigma_\xi^2=100\) to see that they are identical. The kernels will be amplified for higher \(\sigma_\xi^2\), but the shape will not change. 

\section{The Wiener filter is identical to that found by the IBM when \(\gamma\rightarrow0\).}
\label{Wiener}
\begin{figure}
    \centering
    \includegraphics[scale=0.4]{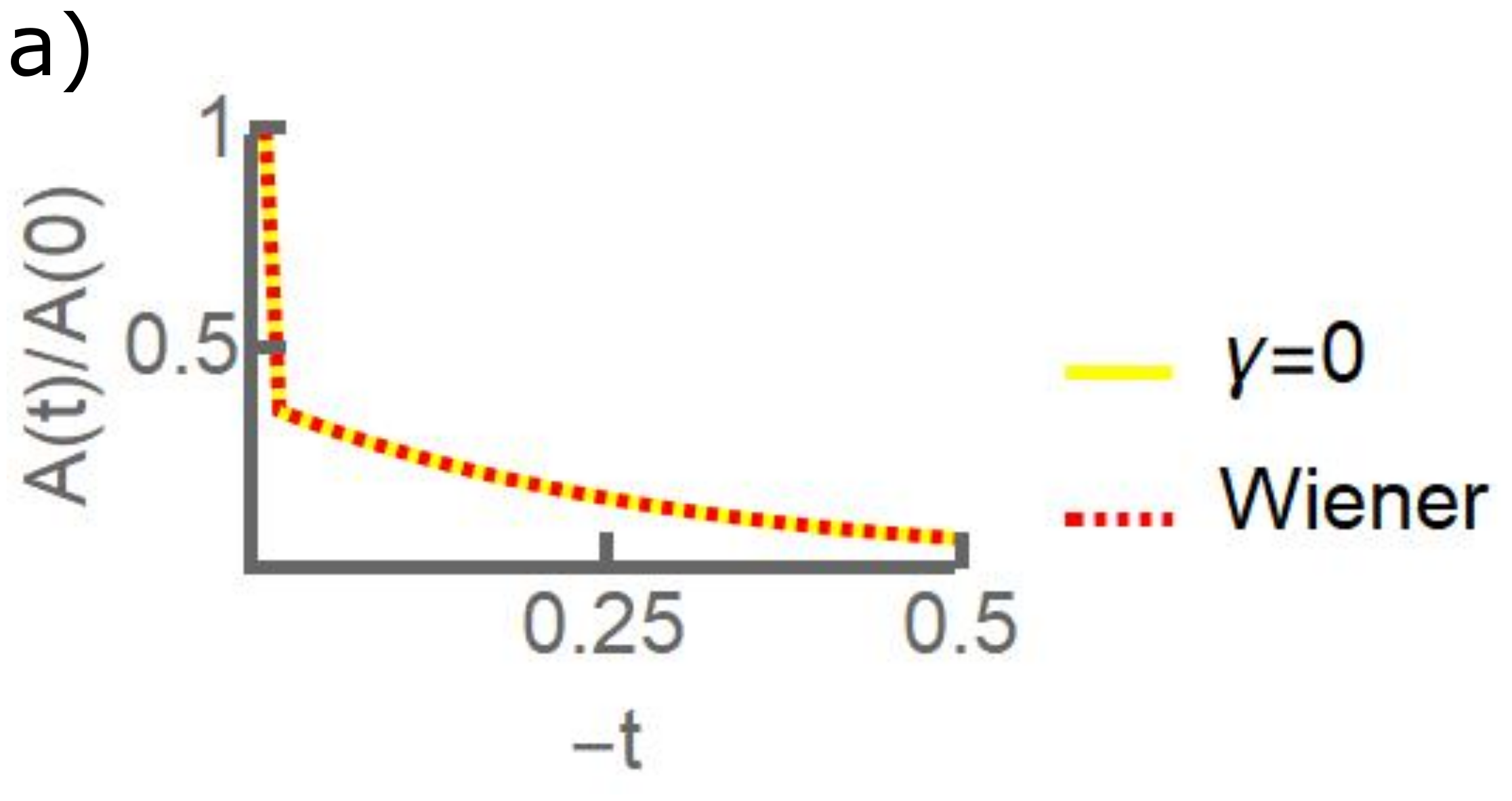}
    \caption{When \(\gamma-0\) the optimal kernel found by the IBM with noise (solid lines) is the same the Wiener filter (dashed line). Thus the technique outlined in the paper can be considered a Wiener filter with compression. Here \(\sigma_{\eta}^2= 2\), \(\sigma=1\), \(\tau_s=2\)s, \(\tau=1\)s, \(\sigma_\xi^2=1\), \(dt=0.01\)s, and \(T=0.5\)s. Note, all kernels have been rescaled so \(A(0)=1\).}
    \label{WienerCompare}
\end{figure}

The discrete Wiener filter minimises the mean squared error between the output \(t\) and the future point \(y\) for a signal with \(\delta x=\vec{\bf{A}}(\delta \vec{\bf{s}}+ \delta \vec{\bf{\eta}})\) and reduces to
\begin{align}
    A&=\Sigma_{\vec{\bf{s}} s(\tau)}(\Sigma_{\vec{\bf{s}}}+\Sigma_{{\vec{\bf{\eta}}}})^{-1}.
\end{align}
The discrete Wiener filter minimises the mean squared error between the filtered signal \(x\) and the value of the signal at some future timepoint \(s(\tau)\) for a signal which has been corrupted by noise \(\delta x=\vec{\bf{A}}(\delta \vec{\bf{s}}+ \delta {\vec{\bf{\eta}}})\).
\begin{align}
    E[\epsilon^2]&=E[(\langle\delta s(\tau)-\vec{\bf{A}}(\langle\delta \vec{\bf{s}}+\langle\delta{\vec{\bf{\eta}}}))(\delta s(\tau) \rangle-\vec{\bf{A}}(\delta \vec{\bf{s}}\rangle+\delta{\vec{\bf{\eta}}}\rangle))]\\
    &=E[(\langle \delta s(\tau) \delta s(\tau) \rangle-2A \langle \delta s(\tau) \delta \vec{\bf{s}} \rangle+ A \langle \delta s \delta s \rangle \vec{\bf{A}}^T+A \langle \delta{\vec{\bf{\eta}}} \delta{\vec{\bf{\eta}}}\rangle \vec{\bf{A}}^T]
\end{align}
We convert \(\Sigma_{xy}=\langle \delta x \delta y \rangle\) and \(\Sigma_{x}=\langle \delta x \delta x \rangle\). Differentiating with respect to A and equating to zero gives
\begin{align}
    \frac{d E[\epsilon^2]}{dA}&=E[-2\Sigma_{s(\tau) {\vec{\bf{s}}}}+ A \Sigma_{\vec{\bf{s}}} +A \Sigma_{\vec{\bf{\eta}}}]=0\\
    A&=2\Sigma_{s(\tau){\vec{\bf{s}}}}(\Sigma_{{\vec{\bf{s}}}}+\Sigma_{{\vec{\bf{\eta}}}})^{-1}.
\end{align}
As shown in figure \ref{WienerCompare}, the kernel obtained using this method has an identical shape to that found using the IBM with noise as \(\gamma\rightarrow0.\)

\section{Finding the limits in which $\Ipred$ is greater when the kernel takes the form $A(t)=A_{\delta}\delta(t)+A_{\rm exp}e^{-\frac{1}{\tau_A} t}$ as opposed to $A(t)=A_{\rm exp}e^{-\frac{1}{\tau_A} t}$}

\begin{figure}
    \centering
    \includegraphics[scale=0.5]{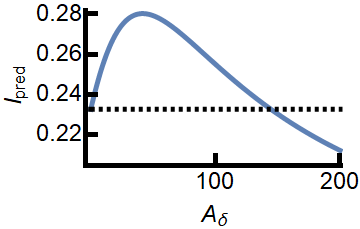}
    \caption{We plot \(A_\delta\) against \(\Ipred\), for \(\sigma_\eta^2=2\), \(\sigma_s^2=1\), \(\tau_s=2\), \(\tau_\eta=0.02\)s, \(\sigma^2_\xi=1\), \(\tau=1\)s. In order to be able to plot it we extract the optimal kernel integration time \(\tau_A^{\rm opt}=0.03476\)s from the discrete system, and set \(A_{\rm exp}=100\) an arbitrarily high value at which the effect of the compression noise \(\xi\) becomes negligible. We see that, like the discrete case (fig. \ref{Optimum}c), the predictive information increases to a peak as \(A_\delta\) increases, giving a finite, non-zero \(A_\delta\) as the optimal value for maximising \(\Ipred\). The dashed line shows \(\Ipred\) when \(A_\delta=0\).}
    \label{deltapeak}
\end{figure}

Consider $\Ipred$ in the continuous form for correlated noise; $\Ipred=\frac{1}{2}\log{\frac{\rho_{s}+\rho_{\eta}+\rho_{\xi}}{\rho_{s}+\rho_{\eta}+\rho_{\xi}-\rho_{s,s(\tau)}}}=\frac{1}{2}\log{\left(1+\frac{\rho_{s,s(\tau)}}{\rho_s+\rho_{\eta}+\rho_{\xi}-\rho_{s,s(\tau)}}\right)}$ where
\begin{align}
    \rho_{s}&=\sigma_s^2 \int^0_{-\infty}\int^0_{-\infty}A(t-s)A(t-s') e^{-\frac{1}{\tau_s}|s-s'|} ds ds',\\
    \rho_{s s(\tau)}&=\sigma_s^2 \left(\int^0_{-\infty}A(t-s) e^{-\frac{1}{\tau_s}|\tau-s|} ds\right)^2,\\
    \rho_\eta&=\sigma_{\eta}^2 \int^0_{-\infty}\int^0_{-\infty}A(t-s)A(t-s') e^{-\frac{1}{\tau_\eta}|s-s'|} ds ds',\\
    \rho_{\xi}&=\sigma_\xi^2.
\end{align}
The optimal kernel shape for this system has the form $A(t)=A_{\delta}\delta(t)+\frac{A_{\rm exp}}{\tau_A}e^{-\frac{1}{\tau_A} t}$. Completing the integrals gives
    \begin{align}    \rho_{s}&=\frac{\frac{A_{\rm exp}^2\sigma_s^2}{\tau_A}+\frac{2A_{\delta}A_{\rm exp}\sigma_s^2}{\tau_A}+A_{\delta}^2\sigma_s^2(\frac{1}{\tau_A}+\frac{1}{\tau_s})}{(\frac{1}{\tau_A}+\frac{1}{\tau_s})},\\
    \rho_{s s(\tau)}&=\frac{\sigma_s^2e^{2\tau/\tau_s}(\frac{A_{\rm exp}}{\tau_A}+A_{\delta}(\frac{1}{\tau_s}+\frac{1}{\tau_A}))^2}{(\frac{1}{\tau_s}+\frac{1}{\tau_A})^2} ,\\
    \rho_{\eta}&=\frac{\frac{A_{\rm exp}^2}{\tau_A}\sigma_{\eta}^2+2A_{\delta}\frac{A_{\rm exp}}{\tau_A}\sigma_{\eta}^2+A_{\delta}^2\sigma_{\eta}^2(\frac{1}{\tau_A}+\frac{1}{\tau_\eta})}{(\frac{1}{\tau_A}+\frac{1}{\tau_\eta})},\\
    \rho_{\xi}&=\sigma_\xi^2.
\end{align}
In figure \ref{deltapeak}, we plot \(\Ipred\) against \(A_\delta\). In order to be able to plot it we extract the optimal kernel integration time \(\tau_A^{\rm opt}=0.03476\)s from the discrete system, and set \(A_{\rm exp}=100\) an arbitrarily high value at which the effect of the compression noise \(\xi\) becomes negligible. We see that, like the discrete case (fig. \ref{Optimum}c), the predictive information increases to a peak as \(A_\delta\) increases, giving a finite, non-zero \(A_\delta\) as the optimal value for maximising \(\Ipred\).

\section{Discretising the push-pull network}

To understand our two systems, we compare the discrete covariance function of the IBM;
\begin{align}
    &\Sigma_{x}=\vec{\bf{A}}\Sigma_{{\vec{\bf{s}}}}\vec{\bf{A}}^T + \vec{\bf{A}}\Sigma_{{\vec{\bf{\eta}}}}\vec{\bf{A}}^T +\Sigma_{\xi},
\end{align}
to the full continuous-time covariances of the push-pull motif from ~\cite{tjalma} to extract the individual covariances. 

The push-pull motif consists of receptors on the surface of a cell that detect the concentration of ligand molecules in the environment \(l\) by binding to them (fig.\ref{pushpullcompare}a). Inside the cell, \(X_T\) output molecules diffuse in and out of contact with the \(R_T\) receptors. Output molecules in contact with bound receptors are activated, using ATP to drive the reaction. These molecules then spontaneously deactivate over time. At steady state \(\phi_l\) receptors are bound and \(\phi_x\) output molecules are activated. The deviations of these quantities from their average are modelled with the linear noise approximation as:
\begin{align}
    \delta \dot{RL} = \gamma \delta l(t)-\frac{\delta RL(t)}{\tau_\eta}+\eta_{RL}(t)\\
    \delta \dot{x^*} = \rho \delta RL(t)-\frac{\delta x^*(t)}{\tau_A}+\eta_{x^*}(t)
\end{align}
The covariances of the concentration of ligand molecules \(\langle \delta_l(t_1) \delta_l(t_2)\rangle\), the receptor-ligand binding noise \(\langle\delta_{\eta_{RL}}(t_1)\delta_{\eta_{RL}}(t_2)\rangle\), and the activation noise \(\langle\delta_{\eta_{x^*}}(t_1)\delta_{\eta{x^*}}(t_2)\rangle\) are given by
\begin{align}
&\langle \delta_l(t_1) \delta_l(t_2)\rangle=\sigma_s^2 e^{-\frac{1}{\tau_s} |t_1-t_2|},\label{lcov}\\
&\langle\delta_{\eta_{RL}}(t_1)\delta_{\eta_{RL}}(t_2)\rangle=2 R_T \phi_l (1-\phi_l)\frac{1}{\tau_\eta}\delta(t_1-t_2),\label{etacov} \\
&\langle\delta_{\eta_{\eta_{x^*}}}(t_1)\delta_{\eta_{\eta_{x^*}}}(t_2)\rangle=2X_T\phi_x(1-\phi_x)\frac{1}{\tau_A}\delta(t_1-t_2),\label{xicov}
\end{align}
where \(\rho=\frac{\phi_x(1-\phi_x)X_T}{\tau_A\phi_l R_T}\), \(\gamma= \frac{\phi_l (1-\phi_l) R_T}{\tau_\eta c} \) are constants related to the push-pull network and \(c\) is the average ligand concentration. The covariances of the number of ligand-bound receptors and activated molecules are then:
\begin{align}
&\langle \delta_{RL}(t_1) \delta_{RL}(t_2)\rangle=\nonumber\\
&\int_{-\infty}^{t_1}\int_{-\infty}^{t_2}\left(\gamma^2 \langle \delta_l(t_1') \delta_l(t_2')\rangle+\langle \delta_{\eta_{RL}}(t_1') \delta_{\eta_{RL}}(t_2')\rangle\right) e^{-\frac{1}{\tau_\eta}(t_1-t_1')} e^{-\frac{1}{\tau_\eta}(t_2-t_2')} dt_2' dt_1'\label{RLcov}\\
&\sigma^2_{x^*}=\nonumber\\
&\int_{-\infty}^{0}\int_{-\infty}^{0}\left(\rho^2 \langle \delta_{RL}(t_1') \delta_{RL}(t_2')\rangle+\langle \delta_{\eta_{x^*}}(t_1') \delta_{\eta_{x^*}}(t_2')\rangle\right) e^{-\frac{1}{\tau_A}(-t_1')} e^{-\frac{1}{\tau_A}(-t_2')} dt_2' dt_1'\label{xvar}
\end{align}
Substituting equations \ref{lcov}-\ref{xicov} into equation \ref{RLcov} gives:
\begin{align}
&\langle \delta_{RL}(t_1) \delta_{RL}(t_2)\rangle=\nonumber\\
&+\int_{-\infty}^{t_1}\int_{-\infty}^{t_2}\gamma^2 \sigma_s^2 e^{-\frac{1}{\tau_s} |t_1'-t_2'|} e^{-\frac{1}{\tau_\eta}(t_1-t_1')} e^{-\frac{1}{\tau_\eta}(t_2-t_2')} dt_2' dt_1'\nonumber\\
&+\int_{-\infty}^{t_1}\int_{-\infty}^{t_2}2 R_T \phi_l (1-\phi_l)\frac{1}{\tau_\eta}\delta(t_1'-t_2') e^{-\frac{1}{\tau_\eta}(t_1-t_1')} e^{-\frac{1}{\tau_\eta}(t_2-t_2')} dt_2' dt_1'.
\end{align}
Completing the integrals and simplifying gives:
\begin{align}
&\langle \delta_{RL}(t_1) \delta_{RL}(t_2)\rangle=\nonumber\\
&\gamma^2 \sigma_s^2\frac{(e^{-\frac{|t_1-t_2|}{\tau_s}}-\frac{\tau_\eta}{\tau_s}e^{-\frac{|t_1-t_2|}{\tau_\eta}})}{\frac{1}{\tau_\eta^2}-\frac{1}{\tau_s^2}}+R_T \phi_l (1-\phi_l) e^{-\frac{|t_1-t_2|}{\tau_\eta}}
\end{align}
\(RL\) is now the signal plus signal noise the system acts on. Plugging this expression into eq. \ref{xvar} gives:
\begin{align}
&\sigma^2_{x^*}=\nonumber\\
&\int_{-\infty}^{0}\int_{-\infty}^{0}\left(\sigma_s^2(e^{-\frac{|t_1'-t_2'|}{\tau_s}}-\frac{\tau_\eta}{\tau_s}e^{-\frac{|t_1'-t_2'|}{\tau_\eta}})+\frac{(\frac{1}{\tau_\eta^2}-\frac{1}{\tau_s^2})}{\gamma^2}R_T \phi_l (1-\phi_l) e^{-\frac{|t_1'-t_2'|}{\tau_\eta}}+\frac{(\frac{1}{\tau_\eta^2}-\frac{1}{\tau_s^2})}{\rho^2\gamma^2}2X_T\phi_x(1-\phi_x)\frac{1}{\tau_A}\delta(t_1'-t_2')\right)\nonumber\\
&\times e^{-\frac{1}{\tau_A}(t_1-t_1')} e^{-\frac{1}{\tau_A}(t_2-t_2')} \frac{\rho^2\gamma^2}{(\frac{1}{\tau_\eta^2}-\frac{1}{\tau_s^2})} dt_2' dt_1'\label{putrhogamma}
\end{align}
we next take a factor of \(\frac{\rho^2\gamma^2}{(\frac{1}{\tau_\eta^2}-\frac{1}{\tau_s^2})}\) outside the integral and substitute \(\rho=\frac{\phi_x(1-\phi_x)X_T}{\tau_A\phi_l R_T}\), \(\gamma= \frac{\phi_l (1-\phi_l) R_T}{\tau_\eta c} \) inside the integrals in eq. \ref{putrhogamma}. Taking out the factor allows us to highlight the relative importance of the signal, signal noise and compression noise. This process gives:
\begin{align}
&\sigma^2_{x^*}=\nonumber\\
&\int_{-\infty}^{t_1}\int_{-\infty}^{t_2}\left(\sigma_s^2(e^{-\frac{|t_1'-t_2'|}{\tau_s}}-\frac{\tau_\eta}{\tau_s}e^{-\frac{|t_1'-t_2'|}{\tau_\eta}})+\frac{\tau_\eta^2 c^2}{\phi_l (1-\phi_l) R_T}(\frac{1}{\tau_\eta^2}-\frac{1}{\tau_s^2}) e^{-\frac{|t_1'-t_2'|}{\tau_\eta}}+\frac{(\frac{1}{\tau_\eta^2}-\frac{1}{\tau_s^2})2\tau_\eta^2 c^2\tau_A}{\phi_x(1-\phi_x)X_T (1-\phi_l)^2 }\delta(t_1'-t_2')\right)\nonumber\\
&\times e^{-\frac{1}{\tau_A}(t_1-t_1')} e^{-\frac{1}{\tau_A}(t_2-t_2')} \frac{\rho^2\gamma^2}{(\frac{1}{\tau_\eta^2}-\frac{1}{\tau_s^2})} dt_2' dt_1'
\end{align}
In order to compare our system to the discrete optimal IBM for the autoregressive signal, we must discretise the system. This way, we can identify the relative importance of the signal, signal noise and compression noise in the discrete case. Discretising the integrals gives:
\begin{align}
&\Sigma^2_{x^*}=\nonumber\\
&\sum_{i=0}^{\frac{T}{\Delta t}}\sum_{j=0}^{\frac{T}{\Delta t}}\left(\sigma_s^2(e^{\frac{-|i-j|\Delta t}{\tau_s}}-\frac{\tau_\eta}{\tau_s}e^{\frac{-|i-j|\Delta t}{\tau_\eta}})\Delta t\Delta t+\frac{\tau_\eta^2 c^2}{\phi_l (1-\phi_l) R_T}(\frac{1}{\tau_\eta^2}-\frac{1}{\tau_s^2}) e^{\frac{-|i-j| \Delta t}{\tau_\eta}}\Delta t\Delta t+\frac{(\frac{1}{\tau_\eta^2}-\frac{1}{\tau_s^2})2\tau_\eta^2 c^2\tau_A}{\phi_x(1-\phi_x)X_T (1-\phi_l)^2 }\delta_{ij}\Delta t\right)\nonumber\\
&\times e^{-\frac{|N-i|\Delta t}{\tau_A}} e^{-\frac{|N-j|\Delta t}{\tau_A}} \frac{\rho^2\gamma^2}{(\frac{1}{\tau_\eta^2}-\frac{1}{\tau_s^2})}
\end{align}
Summing over the Kronecker\(\delta\), \(\delta_{ij}\), gives:
\begin{align}
&\Sigma^2_{x^*}=\nonumber\\
&\sum_{i=0}^{\frac{T}{\Delta t}}\sum_{j=0}^{\frac{T}{\Delta t}}\left(\sigma_s^2(e^{\frac{-|i-j|\Delta t}{\tau_s}}-\frac{\tau_\eta}{\tau_s}e^{\frac{-|i-j|\Delta t}{\tau_\eta}})\Delta t\Delta t+\frac{\tau_\eta^2 c^2}{\phi_l (1-\phi_l) R_T}(\frac{1}{\tau_\eta^2}-\frac{1}{\tau_s^2}) e^{\frac{-|i-j| \Delta t}{\tau_\eta}}\Delta t\Delta t\right) e^{-\frac{|N-i|\Delta t}{\tau_A}} e^{-\frac{|N-j|\Delta t}{\tau_A}} \frac{\rho^2\gamma^2}{(\frac{1}{\tau_\eta^2}-\frac{1}{\tau_s^2})}\nonumber \\
+&\sum_{i=0}^{\frac{T}{\Delta t}}\left(\frac{(\frac{1}{\tau_\eta^2}-\frac{1}{\tau_s^2})2\tau_\eta^2 c^2\tau_A}{\phi_x(1-\phi_x)X_T (1-\phi_l)^2 }\Delta t\right) e^{-\frac{2|N-i|\Delta t}{\tau_A}} \frac{\rho^2\gamma^2}{(\frac{1}{\tau_\eta^2}-\frac{1}{\tau_s^2})}
\end{align}
Next we take the limit \(\tau_s>>\tau_\eta\).
\begin{align}
&\Sigma^2_{x^*}=\nonumber\\
&\sum_{i=0}^{\frac{T}{\Delta t}}\sum_{j=0}^{\frac{T}{\Delta t}}\left(\sigma_s^2(e^{\frac{-|i-j|\Delta t}{\tau_s}})+\frac{c^2}{\phi_l (1-\phi_l) R_T} e^{\frac{-|i-j| \Delta t}{\tau_\eta}}\right)\Delta t\Delta t e^{-\frac{|N-i|\Delta t}{\tau_A}} e^{-\frac{|N-j|\Delta t}{\tau_A}}\frac{\rho^2\gamma^2}{(\frac{1}{\tau_\eta^2}-\frac{1}{\tau_s^2})}\nonumber \\
+&\sum_{i=0}^{\frac{T}{\Delta t}}\frac{2 c^2\tau_A}{\phi_x(1-\phi_x)X_T (1-\phi_l)^2 }\Delta t e^{-\frac{2|N-i|\Delta t}{\tau_A}} \frac{\rho^2\gamma^2}{(\frac{1}{\tau_\eta^2}-\frac{1}{\tau_s^2})}\\
=&\sum_{i=0}^{\frac{T}{\Delta t}}\sum_{j=0}^{\frac{T}{\Delta t}}\left(\sigma_s^2 e^{\frac{-|i-j|\Delta t}{\tau_s}}+\sigma_\eta^2 e^{\frac{-|i-j| \Delta t}{\tau_\eta}}\right) \Delta t\Delta t A(|N-i|\Delta t) A(|N-j|\Delta t)+\sum_{i=0}^{\frac{T}{\Delta t}}\sigma_\xi^2\Delta t A(|N-i|\Delta t)^2 
\end{align}
where 
\begin{align}
    A(t)= \sqrt{\frac{\rho^2\gamma^2}{(\frac{1}{\tau_\eta^2}-\frac{1}{\tau_s^2})}}e^{-\frac{t}{\tau_A}}
\end{align},
\begin{align}
    \sigma_\eta^2=\frac{c^2}{\phi_l (1-\phi_l) R_T}\label{sigmaeta}
\end{align}
and
\begin{align}
    \sigma_\xi^2=\frac{2 c^2\tau_A}{\phi_x(1-\phi_x)X_T (1-\phi_l)^2}
\end{align}. 
Similarly, using the Schurr complement formula:
\begin{align}
    &\Sigma^2_{x^*|s(\tau)}=\nonumber\\
    &\sum_{i=0}^{\frac{T}{\Delta t}}\sum_{j=0}^{\frac{T}{\Delta t}}\left(\sigma_s^2 e^{\frac{-|i-j|\Delta t}{\tau_s}}+\sigma_\eta^2 e^{\frac{-|i-j| \Delta t}{\tau_\eta}}\right) \Delta t\Delta t A(|N-i|\Delta t) A(|N-j|\Delta t)+\sum_{i=0}^{\frac{T}{\Delta t}}\sigma_\xi^2\Delta t A(|N-i|\Delta t)^2\nonumber\\
    &-\sigma_s^2 \left(\sum_{i=0}^{\frac{T}{\Delta t}}A(|N-i| \Delta t) e^{-\frac{1}{\tau_s}|\tau+ |N-i| \Delta t|} \Delta t\right)^2.
\end{align}
Finally, the variance given the signal trajectory is:
\begin{align}
    &\Sigma^2_{x^*|s}=\nonumber\\
    &\sum_{i=0}^{\frac{T}{\Delta t}}\sum_{j=0}^{\frac{T}{\Delta t}}\sigma_\eta^2 e^{\frac{-|i-j| \Delta t}{\tau_\eta}} \Delta t\Delta t A(|N-i|\Delta t) A(|N-j|\Delta t)+\sum_{i=0}^{\frac{T}{\Delta t}}\sigma_\xi^2\Delta t A(|N-i|\Delta t).
\end{align}
Now \(\Ipast=\frac{1}{2}\log{\left(\frac{\Sigma^2_{x^*}}{\Sigma^2_{x^*|s}}\right)}\) and \(\Ipast=\frac{1}{2}\log{\left(\frac{\Sigma^2_{x^*}}{\Sigma^2_{x^*|s(\tau)}}\right)}\).

\section{The optimal integration time of the push-pull network for small \(X_T\)}

\begin{figure}
    \centering
    \includegraphics[scale=0.4]{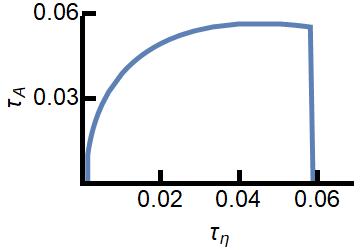}
    \caption{a) {\bf The push-pull kernel initially widens and then sharply narrows as noise correlation time increases (\(\tau_\eta\))}. We plot the correlation time of the kernels for increasing \(\tau_\eta\) found for the push-pull network using a discretised version of the method from~\cite{tjalma} (SI). We see that as \(\tau_\eta\) increases, the width of the optimal kernel initially increases and then decreases sharply, similar to fig. {\ref{TauAndSigma}c}. Here, the average proportion of ligand-bound receptors and activated output molecules is \(\phi_l=\phi_x=0.5\) respectively, and the average concentration of ligands \(c=1\). In all, \(\sigma_s^2=2\times10^{-4}\). In a) and b) \(\tau_\eta=0.02\)s. In b) and c) \(R_T=1\times10^4\) and in a) and c) \(X_T=5\times10^{4}\).}
    \label{SItau}
\end{figure}

In ref. ~\cite{malaguti2021theory}, the authors observe that cells using the push-pull motif respond to the correlation time of the noise increasing by initially increasing the integration time of the kernel. Then as \(\tau_\eta\) approaches \(\tau_s\), time integration averages out the signal as well as the noise, lowering the utility of time integration as a strategy. As \(\tau_\eta\) increases further, the optimal integration time drops sharply back to zero (fig. \ref{SItau}c), mirroring the equivalent drop found for our optimal system (fig. \ref{TauAndSigma}d). Here, the push-pull network's best strategy is merely to capture the current value of the signal, despite noise corruption. 

\end{document}